\documentclass[10pt,draftcls, onecolumn]{IEEEtran}
\usepackage{amsfonts,amsmath,amssymb}
\usepackage{epsfig}
\usepackage{epsfig}

\newtheorem{theorem}{Theorem}

\newtheorem{definition}{Definition}

\newtheorem{remark}{Remark}

\begin{document}

\title{Direct and Indirect Couplings in Coherent Feedback Control of Linear Quantum Systems
\thanks{This work was partially supported by the National Natural Science Foundation of China under Grant 60804015, RGC PolyU 5203/10E, AFOSR Grant FA2386-09-1-4089 AOARD 094089 and Australian Research Council Grant DP0986299.}}

\author{Guofeng~Zhang\thanks{%
G. Zhang is with the School of Electronic Engineering, University of Electronic Science and Technology of China,
Chengdu, Sichuan, China \ 610054. Part of the research was performed when he was with the Department of Applied Mathematics, the Hong Kong Polytechnic University, Hong Kong, China (e-mail: gfzhang@uestc.edu.cn).} ~~ Matthew~R.~James\thanks{M. R. James is with the College of Engineering and Computer Science, Australian National University, Canberra, ACT 0200, Australia (e-mail: Matthew.James@anu.edu.au).}
}
\maketitle

\begin{abstract}
The purpose of this paper is to study and design direct and indirect couplings for use in coherent feedback control of a class of linear quantum stochastic  systems. A general physical model for a nominal linear quantum system  coupled directly and indirectly to external systems is presented. Fundamental properties of stability, dissipation, passivity,  and gain for this class of linear quantum models are presented and characterized using complex Lyapunov equations and linear matrix inequalities (LMIs). Coherent $H^\infty$ and LQG synthesis methods are extended to accommodate direct couplings using multistep optimization. Examples are given to illustrate the results.

\textbf{Index Terms---} quantum feedback control, coherent feedback, dissipation, passivity, gain, $H^{\infty}$ control, LQG control.
\end{abstract}







\section{Introduction}
\label{sec:intro}

Quantum feedback control involves the interconnection of two systems (the quantum plant and the controller) in such a way that the plant-controller system achieves desired performance specifications. The controller may be a quantum system, or   a classical system (i.e. not quantum), or a mixture of the two. When the controller is a classical system, the feedback loop involves measurement, and hence is referred to as {\em measurement feedback}, see \cite{Belavkin83}, \cite{Belavkin92}, \cite{BNM09}, \cite{DJ99}, \cite{JS08},  \cite{KY09},  \cite{MvH07}, \cite{vHSM05}, \cite{WM93},  \cite{WMW02}, \cite{YB09}.  Measurement feedback necessarily involves the loss of quantum coherence in the feedback loop. Measurement feedback control has been applied to the theoretical study of spin localization \cite{Belavkin92}, continuous-time quantum error correction  \cite{CLG08}, \cite{KNPM10}, (non-orthogonal) optical coherent state discrimination \cite{CMG07}, fast state purification \cite{CWJ08}, quantum entanglement generation and preservation \cite{CRH08}, deterministic state preparation \cite{SvHM04}, spin squeezing \cite{TMW02}, \cite{vHSM05b},  atom lasers \cite{YJ09},  feedback control of Bose-Einstein condensates (BECs) \cite{SHCH09}, and so on. Interested readers may refer to  survey papers and books \cite{vHSM05b}, \cite{MK05},  \cite{BvHJ07},  \cite{BvHJ09}, \cite{WM09} for more detailed discussions on measurement feedback control. When the controller is a quantum system, the  controller exchanges quantum information with the quantum plant in the feedback loop. Such quantum information may flow directionally as a (possibly non-commutative) signal like a quantized electromagnetic field or an injected laser \cite{Car93}, \cite{Gar93}, or directly via a bidirectional physical coupling \cite{WM94}, \cite{Lloyd00}. Fully quantum feedback is known as \emph{coherent control}. Compared to measurement feedback control, the benefits of coherent feedback include (i) preservation of quantum coherence within the whole network, and (ii) high speed (a coherent controller would have similar time scale to the plant, and likely is much faster than classical processing).  For convenience, in this paper the directional information flow is called
\emph{indirect coupling}, while the bidirectional coupling is called \emph{direct coupling}.

Several interesting coherent control schemes for quantum systems have been proposed in the  literature. All-optical  (indirect and direct) versus electromechanical measurement feedback schemes are compared in \cite{WM94}, and it is shown that squeezed fields may be produced by coherent  feedback but not by measurement feedback. Further, the authors show in  \cite{WM94} how direct couplings may be used in a feedback loop. Coherent feedback schemes are discussed in \cite{Lloyd00}, e.g. an ion (the plant) in an ion trap can be manipulated desirably by adjoining another ion (the controller) to the trap which interacts with the ion of interest via some appropriately designed  coupling.  A scheme is proposed in \cite{SM06} to produce continuous-wave fields or pulses of polarization-squeezed light by passing classical, linearly polarized laser light through an atomic sample twice; that is, indirect coupling is utilized to feed the field output of the first pass back to the atomic sample again so as to generate polarization-squeezed light. This scheme is confirmed and extended in \cite{SSM08}.  In \cite{Gough08} indirect coupling is employed to construct bilinear control Hamiltonians which have widespread use in quantum applications; bilinear control Hamiltonians are obtained by feeding the output field back to the system so as to cancel out the stochastic effects.

Systematic optimization-based methods for designing indirect coherent feedback systems are  given in \cite{JNP08},  \cite{NJP09} and \cite{MP09}. These methods apply to a class of linear quantum stochastic systems; here the word \lq{linear}\rq \ refers to the linearity of the Heisenberg equations of motion for system operators, while the term \lq{stochastic}\rq \ refers to the quantum noise used to describe the effects of external fields or heat baths interacting with the system.  An algebraic criterion for physical realizability was given in \cite{JNP08}, and this was used in an essential way to determine the coherent quantum controllers in   \cite{JNP08} and \cite{NJP09}. A detailed account of physical realizability is given in \cite{NJD09}. In \cite{JNP08} a  general framework for quantum $H^{\infty }$ control is developed, where a quantum version of bounded real lemma is proposed and  applied to derive necessary and sufficient conditions for the $H^{\infty }$ control of linear quantum stochastic systems.   The paper \cite{Mabuchi08} reports a successful experimental demonstration of this approach. A coherent quantum LQG problem has been addressed in \cite{NJP09}, where it is shown that this problem turns out to be more challenging than the coherent $H^{\infty}$ quantum control in that a property of separation of control and physical realizability does not hold for the coherent LQG problem. The problem of quantum LQG control has also been studied in papers like \cite{DHJMT00}, \cite{WD05} and \cite{DDJW06}, in the framework of measurement-based feedback control.  Linear quantum systems are important not only because of their practical importance, as in quantum optics, but also because of their tractability.

Stability, dissipation, passivity and gain are among the characteristics  fundamental to analysis and synthesis of feedback systems. The paper \cite{JG10} extends J.C.~Willems' theory of dissipative systems to open quantum systems described by quantum noise models. This theory features a physical representation of external systems (or \lq{disturbances}\rq \ or \lq{exosystems}\rq) connected to the system of interest (via direct and/or indirect couplings), and the dissipation inequality is defined in these terms. Storage functions are system observables, such as energy, and may be used as Lyapunov functions to determine stability. The paper \cite{JG10}  also presents an approach to coherent feedback design using physical interconnections.

This paper concerns the influences and uses of  both direct and indirect couplings in coherent quantum feedback system analysis and design. The  specific aims of this paper are as follows. The first aim is to present a general physical model for a nominal linear quantum system  coupled directly and indirectly to external systems. The linear quantum model features explicit algebraic expressions for the dynamical system matrices in terms of physical parameters, as in \cite{GJN10}, and important relations including those required for physical realization. In most of the paper, the dynamical equations are given in the so-called annihilation-creation form \cite[Chapter 7]{WM08}, \cite{GJN10}, which leads to complex system matrices; it is shown how to convert to a real quadrature form similar to that  used in  \cite{JNP08} and \cite{NJP09}. The model provides a deeper physical understanding for the linear quantum models and disturbances studied in  \cite{JNP08}, \cite{NJP09}, \cite{MP09}, \cite{SP09}. The second objective is to describe stability, dissipation, passivity,  and gain for this class of linear quantum models. The definitions and characterizations of these properties are specializations of those given for more general systems in \cite{JG10}. The characterizations are algebraic, involving complex Lyapunov equations and linear matrix inequalities (LMIs). These results generalize recent results for special cases in \cite{JNP08}, \cite{NJP09}, \cite{MP09}, \cite{SP09}. Our final objective is  to extend the $H^\infty$ and LQG design methods to include direct couplings.  At present, no explicit closed-form solutions are known. We give simple examples showing how direct optimization methods may be used. More generally, we present     multi-step optimization schemes  that accommodate direct coupling in coherent $H^\infty$ and LQG design.

This paper is organized as follows. In section 2, the linear quantum systems of interest are introduced. Section II-A presents models of closed (namely, isolated) quantum systems, Section II-B introduces direct coupling between two closed quantum systems, Section II-C discusses indirect coupling mediated by quantum field channels, and Section II-D presents a more general model which is the main concern of this paper. In Section 3, stability and dissipation theory is developed for the above-mentioned linear quantum systems, and versions of the positive real and bounded real lemmas are given. Several examples illustrate these properties and characterization  results. Section \ref{sec:closed_loop} formulates the closed-loop linear quantum systems containing both direct and indirect couplings. An example is given in Section \ref{sec:synth-stab} to demonstrate how direct coupling may be used to stabilize a feedback system. Coherent $H^{\infty }$ and LQG controller synthesis problems  are  investigated in sections \ref{sec:h-infty-synthesis} and \ref{sec:LQG-synthesis} respectively. Examples are given to illustrate the controller design procedures.

Finally, some words for notation.

\textbf{Notation}. Given a column vector of operators $x=[
\begin{array}{ccc}
x_{1} & \cdots & x_{m}%
\end{array}%
]^{T}$ where $m$ is a positive integer, define $x^{\#}=[
\begin{array}{ccc}
x_{1}^{\ast } & \cdots & x_{m}^{\ast }%
\end{array}%
]^{T}$, where the asterisk $\ast $ indicates Hilbert space adjoint or complex conjugation. Furthermore, define the doubled-up column vector to be $%
\breve{x}=[
\begin{array}{cc}
x^{T} & \left( x^{\#}\right) ^{T}%
\end{array}%
]^{T}$. The matrix case can be defined analogously. The symbol ${\rm diag}_{n}\left( M\right) $ is a block diagonal matrix where the square matrix $M$ appears $n$ times as a diagonal block. Given two matrices $U$, $V\in \mathbb{C}^{r\times k}$, a doubled-up matrix $\Delta \left( U,V\right) $ is defined as $\Delta \left( U,V\right) :=[
\begin{array}{cccc}
U & V; & V^{\#} & U^{\#}%
\end{array}%
] $. Let $I_{n}$ be an identity matrix. Define $J_{n}={\rm diag}(I_{n},-I_{n})$ and $\Theta_{n}=[0 ~ ~I_{n}; ~ -I_{n} ~~ 0 ]$. (The subscript ``$n$'' is always omitted.) Then for a matrix $X\in \mathbb{C}^{2n\times 2m}$, define $X^{\flat }:=J_{m}X^{\dagger }J_{n}$.

\section{Linear Quantum Systems} \label{sec:models}

A linear  quantum system $G$ consists of $n$ interacting quantum harmonic oscillators (\cite[Chapter 4]{GZ04})  with annihilation operators $a=[ a_1, \ldots, a_{n} ]^T$ whose evolutions are given by linear differential equations. The annihilation operators are defined by $(a_j \psi)(x) =\frac{1}{\sqrt{2}} x_j \psi(x) +\frac{1}{\sqrt{2}} \frac{\partial  \psi}{\partial x_j}(x)$ on a domain of functions $\psi$ in the Hilbert space $\mathfrak{H}_0= L^2( \mathbb{R}^n, \mathbb{C})$ and satisfy the commutation relations $[a_j, a_k^\ast]=\delta_{jk}$. The nature of the  differential equations satisfied by $a$  depends on whether or not the system is {\em closed}   (isolated from all other systems), or {\em open} (interacting with other systems), but is always defined in terms of unitary evolution $a_j(t) = U^\ast(t) a_j U(t)$, where $U(t)$ is a unitary operator given by Schrodinger's equation. While $a_j$, the initial annihilation operator, is defined on $\mathfrak{H}_0$, at any time $t$ the operator $a_j(t)$ may act on a possibly larger Hilbert space $\mathfrak{H}$ that may include external degrees of freedom. The unitary $U(t)$ is defined on this possibly larger system.

\begin{figure}[tbh]
\epsfxsize=3in
\par
\epsfclipon
\par
\centerline{\epsffile{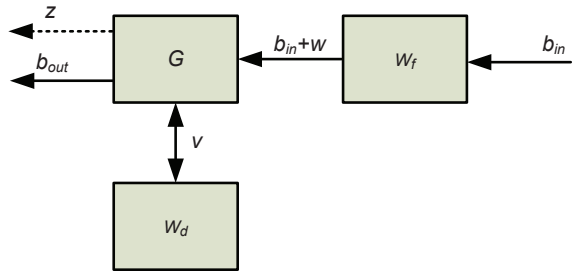}}
\caption{General model}
\label{general_model}
\end{figure}

The general model considered in this paper  is illustrated in Figure \ref{general_model}. This arrangement provides a schematic representation of how a physical system of interest might be influenced by external systems (e.g. uncertainty, disturbances) and noise. The external system $W_d$ is directly coupled to $G$, while the external system $W_f$ is coupled indirectly to $G$ via a quantum field, which is described using a quantum noise signal $b_{in}(t)$. This model is described in detail in  section \ref{sec:models-general}. The intervening sections  explain the physical and system-theoretic basis for the general model by discussing  models for special cases.

In  this paper  we assume that all systems are initialized in a Gaussian state, and that all field inputs are Gaussian \cite[Chapter 4]{GZ04}.

\subsection{Closed Systems} \label{sec:models-closed}

In this section we present a dynamical model of a closed quantum system $G$  of an interconnection of $n$ quantum harmonic oscillators. According to quantum mechanics, the behavior of a closed quantum system $G$  is determined by the Hamiltonian
\begin{equation}
H_0 =\frac{1}{2}\breve{a}^{\dagger }\left[
\begin{array}{cc}
\Omega _{-} & \Omega _{+} \\
\Omega _{+}^{\#} & \Omega _{-}^{\#}%
\end{array}%
\right] \breve{a},
\label{H}
\end{equation}%
where $\Omega _{-}$ and $\Omega _{+}$ are respectively $\mathbb{C}^{n  \times n }$ matrices satisfying $\Omega _{-}=\Omega_{-}^{\dagger }$ and $\Omega _{+}=\Omega _{+}^{T}$. In the Heisenberg picture, the evolution of the annihilation operators is defined by $a_j(t) = U^\ast(t) a_j U(t)$, where $U(t)$ is the solution of Schrodinger's equation
\begin{equation}
\frac{d}{dt} U(t) = -i H_0  U(t), \ \ U(0)=I;
\label{eq:schrodinger-1}
\end{equation}
that is, $\dot a_j (t) =-i [a_j(t) , H_0 (t) ]$,   $a_j(0)=a_j$. This leads to the linear differential equation
\begin{equation}
\dot a(t) = -i \Omega_- a(t) -i \Omega_+ a^\#(t) .
\label{eq:closed-1}
\end{equation}
Note that the equation for $a$ depends on $a^\#$, and so we combine this with the differential equation for $a^\#$ to obtain
\begin{equation}
\left[ \begin{array}{c}
\dot a(t)
\\
\dot a^\#(t)
\end{array}
\right]
= -\left[
\begin{array}{cc}
i\Omega _{-} & i\Omega _{+} \\
-i\Omega _{+}^{\#} & -i\Omega _{-}^{\#}%
\end{array}%
\right] \left[ \begin{array}{c}
 a(t)
\\
a^\#(t)
\end{array}
\right] .
\label{eq:closed-2}
\end{equation}
This may be written compactly in the form
\begin{equation}
\dot{\breve{a}}(t)  = A_0 \breve{a}(t)
\label{eq:closed-3}
\end{equation}
with initial condition $\breve{a}(0)=\breve{a}$, where
\begin{equation}
A_0 = - \Delta( i\Omega_-, i \Omega_+).
\label{eq:A-1}
\end{equation}
We therefore see that the evolution of the  system $G$ is described by the linear differential equation (\ref{eq:closed-3}).

\subsection{Direct Coupling} \label{sec:models-direct}

\begin{figure}[tbh]
\epsfxsize=0.85in
\par
\epsfclipon
\par
\centerline{\epsffile{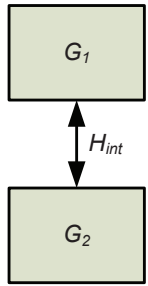}}
\caption{Directly coupled system $G_1 \bowtie G_2$.}
\label{direct_coupling}
\end{figure}

In quantum mechanics, two independent systems $G_1$ and $G_2$ may interact by exchanging energy. This energy exchange may be described by an {\em interaction Hamiltonian} $H_{int}$ of the form $H_{int} = X_1^\dagger X_2 + X_2^\dagger X_1$, where $X_1$ and $X_2$ are respectively  vectors of operators associated with system $G_1$ and $G_2$; see, eg., \cite{WM94}, \cite{Lloyd00}. We will denote the directly coupled system by $G_1 \bowtie G_2$, see Fig.~\ref{direct_coupling}.

We note that the independence of the systems $G_1$ and $G_2$ means that operators associated with $G_1$ commute with those associated with $G_2$.\footnote{Mathematically, if the systems $G_1$ and $G_2$ have Hilbert spaces $\mathfrak{H}_1$ and $\mathfrak{H}_2$ respectively, then the interacting system $G_1 \bowtie G_2$ is a composite system defined on the tensor product space $\mathfrak{H}_1 \otimes \mathfrak{H}_2$.} However, the dynamical evolution of $G_1$ will depend on the evolution of $G_2$, and vice versa.

In what follows we take the interaction Hamiltonian to be
\begin{eqnarray}
H_{int}
=  \frac{1}{2} \left(  \breve{a}^{(1) \dagger} S^\dagger  \breve{a}^{(2)} + \breve{a}^{(2) \dagger} S   \breve{a}^{(1)}
\right),
\label{eq:direct-1}
\end{eqnarray}
where $ S = \Delta( i K_-,i K_+) $ for matrices $K_{-}, K_{+} \in \mathbb{C}^{n_{2}\times n_{1}}$. The Hamiltonian for the directly coupled  system $G_1 \bowtie G_2$ is
\begin{equation}
H = H_{0,1} + H_{int} +H_{0,2} ,
\label{eq:direct-3}
\end{equation}
where $H_{0,k} = \frac{1}{2} \breve{a}^{(k)\dagger} \Delta( \Omega^{(k)}_-, \Omega^{(k)}_+) \breve{a}^{(k)}$ is the self-Hamiltonian for $G_k$, and $H_{int}$ is given by (\ref{eq:direct-1}).

The evolution of the interacting system $G_1 \bowtie G_2$ is determined by Schrodinger's equation $\dot U = -i H U$.
The annihilation operators for $G_1$ evolve according to $a^{(1)}_j(t)= U^\ast(t) a^{(1)}_j U(t)$, which leads to the linear differential equation
\begin{equation}
\dot{\breve{a}}^{(1)}(t)  = A_{0,1} \breve{a}^{(1)} (t) + B_{12} \breve{a}^{(2)}(t) ,
\label{eq:direct-4}
\end{equation}
with initial condition $\breve{a}^{(1)}(0)=\breve{a}^{(1)}$, where $A_{0,1}=-\Delta(i \Omega^{(1)}_-, i \Omega^{(1)}_+)$, and
\begin{equation}
B_{12} = - \Delta( K_-, K_+)^\flat .
\label{eq:direct-5}
\end{equation}
Equation (\ref{eq:direct-4}) concerns the evolution of system $G_1$ influenced by the system $G_2$ via the \lq\lq{input}\rq\rq \ $\breve{v}^{(1)} (t)=   \breve{a}^{(2)} (t) = U^\ast(t) \breve{a}^{(2)} U(t)$. Note that the time trajectory $t \mapsto \breve{v}^{(1)}(t)$ is determined as a result of the interaction between the two systems.

Similarly, we see that the evolution of $G_2$ is given by
$\dot{\breve{a}}^{(2)}(t)  = A_{0,2} \breve{a}^{(2)} (t) + B_{21} \breve{a}^{(1)}(t)$,
with initial condition $\breve{a}^{(2)}(0)=\breve{a}^{(2)}$, where $A_{0,2}=-\Delta(i \Omega^{(2)}_-, i\Omega^{(2)}_+)$, and
\begin{equation}
B_{21} = - B_{12}^\flat = \Delta( K_-, K_+) .
\label{eq:direct-7}
\end{equation}
Of course,  $G_1 \bowtie G_2$ is a special case of the closed systems considered in section \ref{sec:models-closed}, with the interaction terms appearing as off-diagonal terms in the overall Hamiltonian when expressed as a quadratic form in $\breve{a}^{(1)}$ and $\breve{a}^{(2)}$.

\subsection{Indirect Coupling via Quantum Fields} \label{sec:models-indirect}

We consider now a system $G$ coupled to a boson field $\mathcal{F}$.  Boson fields may be used to interconnect component subsystems, effecting {\it indirect  coupling} between them. While the interaction between the system and field may be described from first principles in terms of an interaction Hamiltonian, it is much more convenient to use an idealized quantum noise model which is valid when suitable rotating wave and Markovian conditions are satisfied, as in many situations in quantum optics, eg., cascaded open systems, see \cite{YD84}, \cite{Gar93}, \cite{Car93}, \cite{YK03a} for detail.

The quantum noise models   have a natural input-output structure. The  $m$-channel field before interaction is described by $b_{in}(t)$, the input field, whose components satisfy the singular commutation relations
$$
[ b_{in,j}  (t) , b_{in,k}^\ast(t') ] = \delta_{jk} \delta(t-t'), ~ [ b_{in,j}  (t) , b_{in,k}(t') ]= 0,
$$
$$[ b_{in,j}^\ast(t) , b_{in,k}^\ast(t') ]=0 , ~ (j,k=1, \ldots, m) .
$$
The operators $b_{in,j}(t)$ may be regarded as a quantum stochastic process \cite[Chapter 5]{GZ04}; in the case where the field is in the vacuum state, this process is quantum white noise---a quantum counterpart of Gaussian white noise with zero mean. The integrated processes $B_{in,j}(t) = \int_0^t b_{in,j}(\tau) d\tau$ may be used to define quantum stochastic integrals, with associated non-zero It$\bar{\mbox{o}}$ products $dB_{in,j}(t) dB_{in,k}^\ast(t) = \delta_{jk} dt$. For later use, we define the Ito matrix $F$ by
\begin{equation}
\label{F}
Fdt=(d\breve{%
B}_{in}^{\#}(t)d\breve{B}_{in}^{T}(t))^{T}=
\left[
\begin{array}{cc}
0_{m} & 0  \\
0     &  I_{m}
\end{array}
\right]dt .
\end{equation}

The coupling (interaction) of the system $G$ and the field $\mathcal{F}$ is characterized by the vector $L$ of coupling operators $L = C_- a + C_+ a^\#$ for suitable matrices $C_{-}, C_{+} \in \mathbb{C}^{m \times n}$.  The Schrodinger's equation for the system $G$ (with self-Hamiltonian $H_0$) and field $\mathcal{F}$ is, in It$\bar{\mbox{o}}$ form (\cite[Chapter 11]{GZ04}),
\begin{eqnarray*}
dU(t) = \left\{ L dB^\dagger_{in} - L^\dagger dB_{in} - (\frac{1}{2} L^\dagger L + iH_0 )dt \right\} U(t),
\end{eqnarray*}
with $U(0)=I$. The annihilation operators evolve according to $a_j(t) = U^\ast(t) a_j U(t)$, with It$\bar{\mbox{o}}$ dynamics,
\begin{equation}
d {\breve{a}}(t) = (A_0+A_f) \breve{a}(t)dt  + B_f d\breve{B}_{in}(t),
\label{eq:indirect-2-a}
\end{equation}
where $A_0$ is given by (\ref{eq:A-1}) and $A_f = - \Delta( \Gamma_-, \Gamma_+)$,  $B_f = - \Delta( C_-, C_+)^\flat$,
and
\begin{equation}
\Gamma_\mp = \frac{1}{2} \left(
C_-^\dagger C_\mp - C_+^T C_\pm^\# \right)  .
\label{eq:indirect-4}
\end{equation}
Equation (\ref{eq:indirect-2-a}) may be written in Stratonovich form
\begin{equation}
\dot{\breve{a}}(t) = (A_0+A_f) \breve{a}(t) + B_f \breve{b}_{in}(t) .
\label{eq:indirect-2}
\end{equation}
The output field $b_{out}(t)=U^\ast(t) b_{in}(t) U(t) $ is given by the equation
\begin{equation}
b_{out}(t) = C_f \breve{a}(t) + b_{in}(t),
\label{eq:indirect-5}
\end{equation}
where $ C_f = \Delta(C_-, C_+)$.

\begin{figure}[tbh]
\epsfxsize=3.5in
\par
\epsfclipon
\par
\centerline{\epsffile{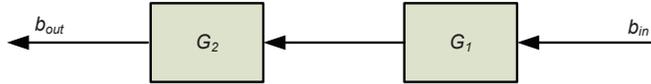}}
\caption{Cascade connection $G_{2}\triangleleft G_{1}$}
\label{cascade}
\end{figure}

Now suppose we have two such systems $G_1$ and $G_2$, specified by self-Hamiltonians $H_0^{(1)}$  and $H_0^{(2)}$ and field coupling operators $L^{(1)}$  and $L^{(2)}$ , respectively.  If the output of $G_1$ is fed into the input of system $G_2$ in a cascade or series connection, the resulting system is denoted $G_2 \triangleleft G_1$; see Fig.~\ref{cascade}. The system $G_2 \triangleleft G_1$ has Hamiltonian $H=H_0^{(1)} + H_0^{(2)} + \mathrm{Im} \{ L^{(2)\dagger} L^{(1)}  \}$ and field coupling operator
$L= L^{(1)} + L^{(2)}$.

In general, the systems $G_1$ and $G_2$ need not be independent, and there may also be  terms describing scattering among field channels (not considered here), \cite{GJ08}. The construction $G_2 \triangleleft G_1$ is called the series product of the two systems, and characterizes  indirect coupling  between the systems via the field channels. A wide range of quantum networks may be described using this and a more general linear fractional transformation (LFT) framework. \cite{YK03a}, \cite{GJ09}.

In this paper we take $G_1$ and $G_2$ as independent systems, comprising $n_1$ and $n_2$ oscillators with annihilation operator vectors $a^{(1)}$ and $a^{(2)}$ respectively.  The system $G_2 \triangleleft G_1$ consists of $n_1+n_2$ oscillators coupled to the field $\mathcal{F}$, with dynamics and output  given by
\begin{eqnarray}
 \dot{\breve{a}}^{(1)}(t)  & \hspace{-2mm}  =& \hspace{-2mm}  (A_{0,1} + A_{f1}) \breve{a}^{(1)}(t) + B_{f1} \breve{b}_{in}(t) ,  \nonumber \\
  \dot{\breve{a}}^{(2)}(t)  & \hspace{-2mm}  =& \hspace{-2mm}  (A_{0,2} + A_{f2}) \breve{a}^{(2)}(t) + B_{f2} (C_{f1} \breve{a}^{(1)}(t)  + \breve{b}_{in}(t)) ,   \nonumber \\
  \breve{b}_{out}(t) & \hspace{-2mm}  =& \hspace{-2mm} C_{f1} \breve{a}^{(1)}(t) +  C_{f2} \breve{a}^{(2)}(t)  + \breve{b}_{in}(t) , \label{eq:indirect-7}
\end{eqnarray}
where all the matrices are determined by the above discussion. Notice the asymmetry in (\ref{eq:indirect-7}), where $a^{(2)}(t)$ has no influence on $a^{(1)}(t)$, in contrast to the symmetry evident in the equations for $G_1 \bowtie G_2$ given in section \ref{sec:models-direct}.

\subsection{A More General Model} \label{sec:models-general}

We now present a more general model for an open system $G$ with specified mechanisms for direct and indirect couplings, based on the ingredients discussed in the preceding sections, Fig.~\ref{general_model}. Our interest is in the influence of external systems on the given system.  The performance characteristics of interest are encoded in a {\em performance variable}\footnote{A performance variable is chosen to capture some aspect of performance, such as an error quantity, and  so may involve external variables, like a reference signal. Performance variables need not have anything to do with the output quantities associated with direct or indirect couplings to other systems.}\, $z$.

The equations for $G$ (including external couplings and performance variables) are
\begin{eqnarray}
\dot{\breve{a}}(t) & \hspace{-2mm}  =& \hspace{-2mm} A \breve{a}(t) + B_d \breve{v}(t) + B_f \breve{w}(t) + B_f \breve{b}_{in}(t), \label{eq:gen-1-dyn} \\
\breve b_{out}(t) & \hspace{-2mm}  =& \hspace{-2mm} C_f \breve{a}(t) +  \breve{w}(t) + \breve{b}_{in}(t),
\label{eq:gen-1-f}  \\
\breve{z}(t) & \hspace{-2mm}  =& \hspace{-2mm} C_p \breve{a}(t) + D_{pd} \breve{v}(t) + D_{pf}  \breve{w}(t),
\label{eq:gen-1-p}
\end{eqnarray}
The complex matrices in (\ref{eq:gen-1-dyn}) and  (\ref{eq:gen-1-f})  are given by
\begin{equation}
A = -\Delta(i\Omega_-, i \Omega_+) - \Delta(\Gamma_-, \Gamma_+), ~ B_d = -\Delta( K_-, K_+)^\flat,
\label{eq:gen-2}
\end{equation}
\begin{equation}
B_f =  -\Delta( C_-, C_+)^\flat, ~ C_f = \Delta(C_-, C_+),
\label{eq:gen-2b}
\end{equation}
where $\Gamma_\pm$ is given by (\ref{eq:indirect-4}). The matrices $A$ and $B_f$  are specified by the parameters $\Omega_\pm$, and $C_\pm$. In equations  (\ref{eq:gen-1-dyn}) and  (\ref{eq:gen-1-f}), ${b}_{in}$ and ${b}_{out}$ are respectively, the input and output fields for $G$. The term $v$ in (\ref{eq:gen-1-dyn})  is an exogenous quantity associated with another (independent) system $W_d$ with which $G$ is directly coupled via the interaction Hamiltonian
\begin{equation}
H_{int}
=  \frac{1}{2} \left(     \breve{a}^{ \dagger} S^\dagger  \breve{v} + \breve{v}^\dagger S   \breve{a}
\right) ,
\label{eq:gen-3}
\end{equation}
where $S=\Delta(i K_-, i K_+)$. The matrices $K_\pm$ determine $B_d$. The term $w$ in (\ref{eq:gen-1-dyn})  is another  exogenous quantity associated with another (independent) system $W_f$ with which $G$ is indirectly coupled through a series connection. Because of the assumed independence, $w$ and $v$ commute with the mode operators $a_j, a_j^\ast$ for $G$. While $v$ and $w$ are arbitrary external variables, the time evolutions $v(t)$ and $w(t)$ are determined by the evolution of the full interacting system. Indeed, $G$ is a subsystem  of the system  $(G_0 \bowtie W_d) \triangleleft W_f$, where $G_0$ is an oscillator coupled to the fields. Equation (\ref{eq:gen-1-dyn}) gives the evolution for the oscillator variables in $G_0$. The matrices $C_p$, $D_{pd}$ and $D_{pf}$ are used to specify the performance variable $z$. The complete system $G$ is specified by the parameters $G = (\Omega_\pm, C_\pm, K_\pm, C_p, D_{pd}, D_{pf})$. Of these, $\Omega_\pm$,  $C_\pm$ and $K_\pm$ are {\em physical} parameters.

\subsection{Quadrature Representation}\label{sec:models-quadrature}

So far, we have used the annihilation and creation operators $a_j, a_j^\ast$ to represent oscillators systems, using the doubled-up notation $\breve{a} = [a^T ~ a^\dagger]^T$. This {\em annihilation-creation representation} results in equations with complex matrices. There is an alternative {\em quadrature representation}, which  results in equations with real matrices; these may be more convenient for standard matrix analysis software packages.

Define the unitary matrix
\begin{equation}
\Lambda  = \frac{1}{\sqrt{2}} \left[
\begin{array}{cc}
I & I
\\
-i I & i I
\end{array}
\right]
\label{eq:quad-1}
\end{equation}
and the vector
\begin{equation}
\tilde  a  =   \left[
\begin{array}{c}
q
\\
p
\end{array}
\right]
\label{eq:quad-2}
\end{equation}
of self-adjoint operators
by the relation
\begin{equation}
\tilde a = \Lambda  \breve{a}.
\label{eq:quad-3}
\end{equation}
 The vector $q = \frac{1}{\sqrt{2}} [I ~ I] \breve{a}$ is known as the {\em real quadrature}, while $p = \frac{1}{\sqrt{2}} [-i I ~ i I] \breve{a}$ is called the {\em imaginary} or {\em phase quadrature}. Since $\Lambda$ is unitary, equation (\ref{eq:quad-3}) is invertible: $\breve{a} = \Lambda^\dagger \tilde a$.

Now let $\Lambda_f$, $\Lambda_d$ and $\Lambda_p$ be unitary matrices, of suitable dimension,  of the form (\ref{eq:quad-1}), and define the following quadrature vectors $
\tilde b_{in} = \Lambda_f  \breve{b}_{in}$, $\tilde b_{out} = \Lambda_f  \breve{b}_{out}$,
$\tilde w = \Lambda_f  \breve{w}$, $\tilde v = \Lambda_d  \breve{v}$,  $\tilde z = \Lambda_p \breve z$.
Then in quadrature form $G$ is given by
\begin{eqnarray*}
\dot{\tilde{a}}(t) &=&  \tilde  A \tilde{a}(t) +  \tilde  B_d \tilde{v}(t) +  \tilde  B_f \tilde{w}(t) + \tilde  B_f \tilde{b}_{in}(t), \\
\tilde{b}_{out}(t) &=&
 \tilde  C_f \tilde{a}(t) +  \tilde{w}(t) + \tilde{b}_{in}(t),  \\
\tilde{z}(t) &=&  \tilde  C_p \tilde{a}(t) + \tilde  D_{pd} \tilde{v}(t) + \tilde D_{pf}  \tilde{w}(t),
\end{eqnarray*}
where $\tilde A = \Lambda A \Lambda^\dagger$,
 $\tilde B_d  = \Lambda B_d  \Lambda^\dagger_d$,
$\tilde B_f  = \Lambda B_f  \Lambda^\dagger_f$,
 $\tilde C_f  = \Lambda_f C_f  \Lambda^\dagger$,
$\tilde C_p  = \Lambda_p C_p  \Lambda^\dagger$,
$\tilde D_{pd}  = \Lambda_p D_{pd}   \Lambda^\dagger_d$,
$\tilde D_{pf}  = \Lambda_p D_{pf}   \Lambda^\dagger_f$.
 Note that all entries of the matrices in this representation are real.

\subsection{Physical Realization}\label{sec:models-realization}

The matrices in equation  (\ref{eq:gen-2})-(\ref{eq:gen-2b}) for the general linear quantum system $G$ of section \ref{sec:models-general} have special structure, and therefore form a subclass of the class of linear dynamical models. When designing a quantum linear system, as in coherent feedback design, it is important to know when a linear dynamical system corresponds to a quantum linear stochastic systems. This is a fundamental question of {\em physical realization}, \cite{JNP08}, \cite{NJP09}, \cite{NJD09}, \cite{SP09}. In this section we present an additional relation that characterizes the preservation of commutation relations ($[a_j(t), a^\ast_k(t) ] = \delta_{jk}$ for all $t$) which is fundamental to physical realizability.

In the doubled-up notation, the commutation relations may be expressed in the form
\begin{equation}
[ \breve{a}_j, \breve{a}_k^\ast ] = (J_n)_{jk} .
\label{eq:c-0}
\end{equation}
Now it is straightforward to verify that for the matrices $A$ and $C_f$ defined in section \ref{sec:models-general} we have
\begin{equation}
A + A^\flat + C^\flat_f C_f =0.
\label{eq:c-3}
\end{equation}
By multiplying (\ref{eq:c-3}) by $J_n$ on the left, and using  the above definitions we see that the following fundamental relations for the matrices in the general model $G$ of section \ref{sec:models-general} hold:
\begin{eqnarray}
J_n A + A^\dagger  J_n  + C_f^\dagger    J_{m} C_f &=& 0 , \label{eq:c-2-ccr} \\
B_f &=& -C^\flat_f , \label{eq:c-2-f} \\
B_d &=& -\Delta( K_-, K_+)^\flat . \label{eq:c-2-d}
\end{eqnarray}
Equation (\ref{eq:c-2-ccr}) characterizes preservation of the commutation relations, and moreover, the relations
(\ref{eq:c-2-ccr})-(\ref{eq:c-2-d}) generalize the
 canonical physical realizability criteria in \cite[Theorem 3.4]{JNP08}, \cite[Theorem 5.1]{MP09}, \cite[Theorem 3]{SP09}.

\section{Performance Characteristics of Systems with Direct and Indirect Interactions}\label{sec:analysis}

The purpose of this section is to discuss basic performance characteristics such as stability, passivity, gain, etc. for linear quantum systems, using the models developed in section \ref{sec:models}.


\subsection{Stability} \label{sec:stab}

Perhaps the most basic performance characteristic is stability. In the case of a system $G$ based on quantum harmonic oscillators, stability may be assessed in terms of the behavior of the number of quanta (e.g. photons) stored in the system, ${\bf N}=a^{\dagger }a=\sum_{j=1}^{n}a_{j}^{\ast }a_{j}$. With zero inputs but possibly coupled to vacuum fields, that is, $w=0$ and $v=0$ in (\ref{eq:gen-1-dyn}), we will say that $G$ is \emph{exponentially stable} if $\langle {\bf N}(t)\rangle  \leq c_0 e^{-c_1 t } \langle {\bf N} \rangle + c_2$ for some $c_0>0, c_1>0$, and $c_2 \geq 0$, (ii) \emph{marginally stable} if $\langle {\bf N}(t)\rangle \leq c_1 \langle {\bf N} \rangle + c_2 t$ for some $c_1>0$ and $c_2 \geq 0$,   and (iii) \emph{exponentially unstable} if there exists an initial system state such that  $\langle {\bf N}(t)\rangle  \geq c_0 e^{c_1 t } \langle {\bf N} \rangle +c_2$ for some real constant $c_0>0, c_1>0$ and $c_2$.

For example, for the closed systems studied in Sec. \ref{sec:models-closed}, if $\Omega_+=0$ we see that $a(t) = \exp (-i \Omega_- t) a$, and $a^\dagger(t) a(t)=a^\dagger a$ for all $t$, which means that $G$ is marginally stable but not exponentially  stable---it oscillates---hence the name \lq\lq{oscillator}\rq\rq.

The total number of quanta ${\bf N} = a^\dagger a$ is a natural Lyapunov function for $G$, and is directly related to the energy of the system. In what follows we find it is more convenient to use
\begin{equation}
V  = \frac{1}{2} \breve{a}^\dagger \breve{a} = a^\dagger a + \frac{n}{2},
\label{eq:V}
\end{equation}
which differs from the total number of quanta by an additive constant. More generally, we will consider storage functions of the form $V=\frac{1}{2} \breve{a}^\dagger P \breve{a}$ for non-negative Hermitian matrices $P$.

The following result is a simple  criterion for stability of system $G$ defined in section \ref{sec:models-general}.

\begin{theorem}\label{thm:stability}
If there exist $P\geq 0$ and $Q\geq cP$ for a scalar $c>0$ such
that
\begin{equation}
A^{\dagger }P+PA+Q\leq 0,
\label{PQ}
\end{equation}%
then
\begin{equation}
\left\langle \breve a^\dagger(t) P \breve a(t)
\right\rangle \leq e^{-ct}\left\langle \breve a^\dagger P \breve a
\right\rangle
+\frac{\lambda}{2c} ,
\label{V}
\end{equation}%
where $\lambda =\mathrm{tr}[B_{f}^{\dagger }PB_{f}F]$,  and $F$ is defined by (\ref{F}). If also $P \geq  \alpha I$ ($\alpha > 0$), then we have
$\left\langle a^\dagger(t) a(t)
\right\rangle \leq \frac{1}{\alpha}  e^{-ct}\left\langle \breve a^\dagger P \breve a
\right\rangle
+\frac{\lambda}{2c\alpha} .
$
\end{theorem}
The proof of Theorem \ref{thm:stability} is similar to that in \cite{JG10}, so is omitted.

As an example, for the closed systems studied in Sec. \ref{sec:models-closed}, in the special case of $n=1$, $\Omega_-=0$, $\Omega_+=i\epsilon/2$, with $\epsilon$ real, the rate of change of energy is $\frac{\epsilon}{2}((a^\ast)^2 + a^2)$ and the matrix $\frac{1}{2}(A_0 +A_0^\dagger)$ has eigenvalues $\pm \epsilon$. Also, $\frac{1}{2} \Lambda (A_0 +A_0^\dagger)\Lambda^\dagger = \mathrm{diag}(\epsilon, -\epsilon)$, which means that the real quadrature expands, and the phase quadrature contracts---this is the basic \lq\lq{squeezing}\rq\rq \ action, as in a degenerate parametric amplifier (a realistic model would also include damping, see section \ref{sec:analysis-eg-1-dpa}). Therefore the stability of the closed system $G$ depends on $\Omega_+$.

\subsection{Dissipativity of Linear Quantum Systems}\label{sec:analysis-dissipation}

We now consider the general class of open systems  $G$ defined in section \ref{sec:models-general}. In order to simplify the notation we write $u = [v^T \, \, w^T]^T$ for the doubled-up vector of external variables, and define accordingly
\[
B:=[ B_d \, \, B_f]\left[
\begin{array}{cccc}
I & 0 & 0 & 0 \\
0 & 0 & I & 0 \\
0 & I & 0 & 0 \\
0 & 0 & 0 & I%
\end{array}%
\right],
\]
where dimensions of identity matrices are implicitly assumed to be conformal to those of $v$ and $w$. The dynamical equation (\ref{eq:gen-1-dyn})  becomes
\begin{equation}
\dot{\breve{a}}(t) =  A \breve{a}(t) + B  \breve{u}(t)  +  B_f \breve{b}_{in}(t).
\label{eq:diss-1}
\end{equation}
 In order to define dissipation for $G$, we  use the supply rate
 \begin{equation}
r(\breve{a}, \breve{u}) =\frac{1}{2}  [ \breve{a}^\dagger \, \breve{u}^\dagger ] R \left[  \begin{array}{c}
\breve{a} \\ \breve{u} \end{array} \right],
\label{eq:diss-6}
\end{equation}
 where $R$ is a Hermitian matrix of the form
\begin{equation}
 R = \left[ \begin{array}{cc}
 R_{11} & R_{12}
 \\
 R_{12}^\dagger & R_{22}
 \end{array} \right] .
\label{eq:diss-5}
\end{equation}

Given a non-negative Hermitian matrix $P$, we define a candidate storage function $V=  \frac{1}{2} \breve{a}^\dagger P \breve{a}$. In what follows we will need the relation
\begin{eqnarray}
&&\mathbb{E}_{s}[V(t)]  \label{eq:diss-6a} \\
&=&V(s)+\frac{1}{2}\int_{s}^{t}\mathbb{E}_{s}\left[ \breve{a}^{\dagger
}(\tau )(PA+A^{\dagger }P)\breve{a}(\tau )\right. \nonumber  \\
&&+\breve{u}^{\dagger }(\tau )B^{\dagger }P\breve{a}(\tau )+\breve{a}%
^{\dagger }(\tau )PB\breve{u}(\tau )+\mathrm{tr}[B_{f}^{\dagger
}PB_{f}F]]d\tau   \nonumber
\end{eqnarray}%
for $s \leq t$, where the matrix $F$ is that defined in (\ref{F}), and the notation $\mathbb{E}_s$ is used to denote the operation of averaging out the quantum noise in the field channels from time $s$ onwards, c.f. \cite[page 215]{Pa92}.

\begin{definition}  \label{dfn:diss} (Dissipation)
We say that the system $G$ is {\em dissipative} with respect to the supply rate $r(\breve{a}, \breve{u})$ (given by (\ref{eq:diss-6})) if there exists a non-negative Hermitian matrix $P$ and a non-negative real number $\lambda$  such that for
$
V = \frac{1}{2} \breve{a}^\dagger P \breve{a}
$
we have
\begin{equation}
 \mathbb{E}_0 [ V(t) ] \leq   V   + \mathbb{E}_0 [ \int_0^t  r(\breve{a}(\tau), \breve{u}(\tau)) d\tau ] + \lambda t
\label{eq:diss-10}
\end{equation}
for all $t \geq 0$ and  all external variables $u$ and all Gaussian states for the systems and fields.  (Recall that $V(0)=V$ defined in Eq. (\ref{eq:V}).)
\end{definition}


As in \cite[Theorem 4.2]{JNP08}, the dissipativity of $G$ may be characterized in terms of a linear matrix inequality(LMI).

\begin{theorem}   \label{thm:diss} (Dissipation)
The system $G$ is   dissipative with respect to the supply rate $r(\breve{a}, \breve{u})$ (given by (\ref{eq:diss-6}))  if and only if there exists a non-negative Hermitian matrix $P$ such that
\begin{equation}
\left[  \begin{array}{cc}
PA + A^\dagger P - R_{11}  & PB -R_{12}
\\
B^\dagger P - R_{12}^\dagger & - R_{22}
\end{array} \right ] \leq 0.
\label{eq:diss-11}
\end{equation}
Moreover, $\lambda = \mathrm{tr}[ B_f ^\dagger P B_f F ] $.
\end{theorem}

The proof of this Theorem is similar to that of  \cite[Theorem 4.2]{JNP08} and will not be given, except to say that it involves combining  the relation (\ref{eq:diss-6a}) and the inequality (\ref{eq:diss-10}).

\subsection{Positive Real Lemma}
\label{sec:analysis-prl}

In this section, we study passivity for linear quantum systems presented in Sec. \ref{sec:models-general}. Taking  $R_{11}=-Q$, $R_{12}=C_p^\dagger$, $R_{22}=0$, and a performance variable $\breve{z} = C_p \breve{a}$, the supply rate $r$ in (\ref{eq:diss-6}) becomes
\begin{equation}
r(\breve{a}, \breve{u}) = \frac{1}{2}( - \breve{a}^\dagger Q \breve{a} + \breve{u}^\dagger \breve{z} + \breve{z}^\dagger \breve{u})
\label{eq:prl-4}
\end{equation}

\begin{definition}  \label{dfn:passive} (Passivity)
We say that a system $G$ of the form (\ref{eq:diss-1})  with a performance variable $z$ is {\em passive} if and only if it is dissipative with respect to the supply rate defined in (\ref{eq:prl-4}) with $Q$ being non-negative.
\end{definition}

Passivity can be checked using Theorem \ref{thm:diss}. In fact by Theorem \ref{thm:diss} we have the following version of the positive real lemma.

\begin{theorem}  \label{thm:PRL} (Positive Real Lemma)
The system $G$ with performance variable $\breve{z} = C_p \breve{a}$  is passive if and only if there exist non-negative definite Hermitian matrices $P$ and $Q$ such that
\begin{equation}
\left[  \begin{array}{cc}
PA + A^\dagger P + Q  & PB -C_p^\dagger
\\
B^\dagger P - C_p  & 0
\end{array} \right ] \leq 0.
\label{eq:prl-111}
\end{equation}
Moreover, $\lambda = \mathrm{tr}[ B_f^\dagger PB_f F] $.
\end{theorem}

We now discuss  the natural passivity property for the system (\ref{eq:diss-1}), along the lines of \cite[section III.A.]{JG10}. Let $V=\frac{1}{2} \breve{a}^\dagger \breve{a}$, and  note that the
 LMI (\ref{eq:prl-111}) from Theorem \ref{thm:PRL} is satisfied with equality when $P=I$, $C_p = B^\dagger$, and
\begin{equation}
Q = -(A+A^\dagger) = \Delta( C_-^+ C_- - C_+^T C_+^\#, 2i \Omega_+) ,
\label{eq:prl-1}
\end{equation}
which is not necessarily non-negative. Consequently, passivity with respect to the performance variable $\breve{z} = B^\dagger \breve{a}$ will depend  on the definiteness of  $Q$. Interestingly, define $M= M_- a + M_+ a^\# $, where
\begin{equation}
 M_\pm  = \left[  \begin{array}{c}
K_\pm
\\
C_\pm
\end{array}
\right]  .
\label{eq:prl-3}
\end{equation}
Then it is easy to show that the performance variable $z$ can be written as the commutator $z=[V, M]$.

\begin{remark}   \label{rmk:prl-1}
{\rm
In the special case of no direct coupling, with $u=w$, $B=B_f$, we have $C_p=B_f^\dagger$, whereas $C_f = -B_f^\flat$. In general, $C_p \neq C_f$, which means that the performance variable $z$ does not form part of the field output.  However, when $B$ satisfies the invariance condition $B = JBJ$, namely $C_+ =0$, we have $C_p=-C_f$ and
 $\breve{b}_{out}(t) = -\breve{z}(t) + \breve{w}(t)+\breve{b}_{in}(t)$.}
$\Box$
\end{remark}

\subsection{Bounded Real Lemma}\label{sec:analysis-brl}

Theorem \ref{thm:diss} may also be used to characterize the $L^2$ gain property, as discussed in \cite{JNP08},  \cite{MP09}, \cite{SP09}. Here we summarize the main points using the framework developed in this paper.  Our interest is in the influence of the external systems $W_d$ and $W_f$ on the performance variable $z$,  given in the general model by equation (\ref{eq:gen-1-p}), in the sense of $L^2$ gain.

Using the notation $D_p = [ D_{pd} \, \, D_{pf}]$, the performance variable is given by $\breve{z} = C_p \breve{a} + D_p \breve{u}$. The dynamical evolution is given by (\ref{eq:diss-1}). Define a supply rate
\begin{equation}
r(\breve{a}, \breve{u}) = - \frac{1}{2} ( \breve{z}^\dagger \breve{z} - g^2 \breve{u}^\dagger \breve{u} ) ,
\label{eq:brl-1}
\end{equation}
where $g \geq 0$ is a real gain parameter. This corresponds to the choice
$R_{11}= -C_p^\dagger C_p$, $R_{12}=-C_p^\dagger D_p$, $R_{22}=g^2-D_p^\dagger D_p$.

\begin{definition}  \label{dfn:L2gain} (Bounded realness)
For the system $G$ of the form (\ref{eq:diss-1})  with performance variable $\breve{z} = C_p \breve{a}+D_p \breve{u}$, we say the transfer $\breve{u} \mapsto \breve{z}$ is bounded real with finite $L^2$ gain less than $g$ if the system is dissipative with respect to the supply rate defined in (\ref{eq:brl-1}). If the gain inequality holds strictly,  then we say  the transfer $\breve{u} \mapsto \breve{z}$ is  strictly bounded real with disturbance attenuation $g$.
\end{definition}

From  Theorem \ref{thm:diss},  we have:

\begin{theorem}  \label{thm:BRL} (Bounded  Real Lemma)
The system $G$ with performance variable $\breve{z} = C_p \breve{a}+D_p \breve u$  is bounded real with finite $L^2$ gain less than $g$ if and only if there exists a non-negative Hermitian matrix $P$ such that
\begin{equation}
\left[  \begin{array}{cc}
PA + A^\dagger P +C_p^\dagger C_p   & PB  +C_p^\dagger D_p
\\
B^\dagger  P + D_p^\dagger C_p &  D_p^\dagger D_p-g^2 I
\end{array} \right ] \leq 0.
\label{eq:brl-3}
\end{equation}
 Moreover, $\lambda = \mathrm{tr}[ B_f^\dagger PB_f F] $.
\end{theorem}

In the special case $C_+=0$, $\Omega_+=0$, no direct coupling,  $C_p=C_f$, and $D_p=I$, we have $\breve{z} = C_p \breve{a}+\breve{w}$ and $\breve{b}_{out} = \breve{z} + \breve{b}_{in}$. In this case,  the LMI (\ref{eq:brl-3})  is satisfied with equality for $P=I$ and $g=1$, which means that the gain of the transfer $\breve{w} \mapsto \breve{z}$ is {\em exactly} one for such systems. This corresponds to the lossless bounded real property discussed in \cite{MP09}, and shown to be equivalent to physical realizability for this special class of systems. In general, however, physical realizability and the lossless bounded real properties are distinct (see, e.g. section \ref{sec:analysis-eg-1-dpa}).

As in \cite{JNP08}, the following version of the strict bounded real lemma may be proven.

\begin{theorem} (Strict Bounded Real Lemma)
\label{thm:brl} The following statements are equivalent.

\begin{description}
\item[i)] The quantum system $G$ defined in (\ref{eq:gen-1-dyn})-(\ref{eq:gen-1-p}) is strictly bounded real with disturbance attenuation $g$.

\item[ii)] $A$ is stable and $\left\Vert C_{p}\left( sI-A\right)^{-1}B+D_{p}\right\Vert _{\infty }<g$.

\item[iii)] $g^{2}I-D_{p}^{\dagger }D_{p}>0$ and there exists a Hermitian matrix $P_{1}>0$ satisfying inequality%
\begin{equation} \label{LMI_form1}
\left[
\begin{array}{cc}
A^{\dagger }P_{1}+P_{1}A+C_{p}^{\dagger }C_{p} & P_{1}B+C_{p}^{\dagger }D_{p}
\\
B^{\dagger }P_{1}+D_{p}^{\dagger }C_{p} & D_{p}^{\dagger }D_{p}-g^{2}I%
\end{array}%
\right] <0,
\end{equation}%
or, equivalently,%
\begin{equation}
\left[
\begin{array}{ccc}
A^{\dagger }P_{1}+P_{1}A & P_{1}B & C_{p}^{\dagger } \\
B^{\dagger }P_{1} & -gI & D_{p}^{\dagger } \\
C_{p} & D_{p} & -gI%
\end{array}%
\right] <0.  \label{LMI_ori}
\end{equation}

\item[iv)] $g^{2}I-D_{p}^{\dagger }D_{p}>0$ and there exists a Hermitian matrix $P_{2}>0$ satisfying the algebraic Riccati equation
\begin{eqnarray*}
&&A^{\dagger }P_{2}+P_{2}A +\left( P_{2}B+C_{p}^{\dagger
}D_{p}\right)   \\
 && \times\left( g^{2}I-D_{p}^{\dagger }D_{p}\right) ^{-1}(
B^{\dagger }P_{2}^{\dagger }+D_{p}^{\dagger }C_{p})  \\
&=&0
\end{eqnarray*}
with $A+BB^{\dagger }P_{2}$ being Hurwitz.
\end{description}

Furthermore, if these statements hold, then $P_1 < P_2$.
\end{theorem}

\subsection{LQG Performance} \label{sec:analysis-lqg}

In this section a quantum LQG cost function is defined in the annihilation-creation form, and whose evaluation is connected to a Lyapunov equation in the complex domain.

Consider the following quantum system
\begin{equation}
d\breve{a}(t)=A\breve{a}(t)dt+B_{f}d\breve{B}_{in}(t)  \label{system}
\end{equation}%
where $B_{in}(t)$ is a quantum Wiener process introduced in section \ref{sec:models-indirect}. Given a performance variable  $\breve{z}(t)=C_{p}\breve{a}(t)$, define a finite-horizon LQG cost function to be $
\mathfrak{J}(t_{f})=\int_{0}^{t_{f}}\langle \breve{z}^{\dagger }(t)%
\breve{z}(t)\rangle dt $ for arbitrary $t_{f}>0$. Along the line of \cite{NJP09},  the infinite-horizon LQG cost is
\begin{eqnarray}
\mathfrak{J}_{\infty} &=&\lim_{t_{f}\rightarrow \infty }\frac{1}{t_{f}}%
\int_{0}^{t_{f}}\frac{1}{2}\left\langle \breve{z}^{\dagger }(t)%
\breve{z}(t)+\breve{z}^{T}(t)\breve{z}^{\#}(t)\right\rangle dt \nonumber \\
&=&\lim_{t_{f}\rightarrow \infty }\frac{1}{t_{f}}\int_{0}^{t_{f}}\mbox{Tr}\left\{
C_{p}P_{LQG}(t)C_{p}^{\dagger }\right\} dt  \nonumber \\
&=&\mbox{Tr}\left\{
C_{p}P_{LQG}C_{p}^{\dagger }\right\} ,
\label{LQG_index}
\end{eqnarray}
where the constant Hermitian matrix $P_{LQG}$ satisfies the following Lyapunov equation%
\begin{equation}
AP_{LQG}+P_{LQG}A^{\dagger }+\frac{1}{2}B_{f}B_{f}^{\dagger }=0.
\label{lyap}
\end{equation}

\subsection{Examples}\label{sec:analysis-eg-1}

In this section we use several examples to illustrate the stability, passivity and finite $L^2$ gain properties discussed in the preceding sections.

\subsubsection{Oscillator with Directly-Coupled Disturbance}
\label{sec:analysis-eg-0-closed-osc}

Let the system $G$ be an oscillator (no field connections) directly coupled to an external system $W_d$ via an interaction Hamiltonian $H_{int}=i \gamma( v^\ast a - a^\ast v)$. That is, $\Omega_-=\omega$, $\Omega_+=0$. $C_-=0$, $C_+=0$, $K_-=\gamma$, $K_+=0$. This  system evolves according to $\dot a = -i\omega a -\gamma v$, so that for $V=a^\ast a$ we have $\dot V = -\gamma v^\ast a - \gamma a^\ast v$. Hence $G$ is passive with respect to the performance variable $z=-\gamma a$.

\subsubsection{Optical Cavity}\label{sec:analysis-eg-1-cavity}

An optical cavity $G$ is a single open oscillator \cite{GJN10} with $\Omega_-=\omega$, $\Omega_+=0$, $C_-=\sqrt{\kappa}$, $C_+=0$, that is (\cite[section IV. B.]{GJN10}),
$$
\dot{\breve{a}} =\left[
\begin{array}{cc}
-\frac{\kappa }{2}-i\omega  & 0 \\
0 & -\frac{\kappa }{2}+i\omega
\end{array}%
\right] \breve{a}-
\sqrt{\kappa} \,
  \left( \breve{w}+\breve{b}_{in}\right) ,
$$
where $\breve{w}$ arises from indirect coupling to  an external system $W_f$ via the input field.

As in section \ref{sec:analysis-brl}, choose a performance variable $\breve{z} =\sqrt{\kappa } \breve{a}+\breve{w}$. According to the discussion in section \ref{sec:analysis-brl}, $G$ has $L^2$ gain 1. Moreover,  since $
Q=\Delta \left( C_{-}^{\dagger }C_{-},2i\Omega _{+}\right)  =[\kappa~  0;0~ \kappa ]>0$, following the development in section \ref{sec:analysis-prl}, $G$ is exponentially stable and passive with respect to the performance variable $\breve{z} = -\sqrt{\kappa } \breve{a}$.

\subsubsection{Degenerate Parametric Amplifier} \label{sec:analysis-eg-1-dpa}

A degenerate parametric amplifier (DPA) is an open oscillator that is able to produce squeezed output field. A model of a DPA is  as follows (\cite[page 220]{GZ04}):
\begin{eqnarray*}
\dot{\breve{a}} &=&-\frac{1}{2}\left[
\begin{array}{cc}
\kappa  & -\epsilon  \\
-\epsilon  & \kappa
\end{array}%
\right] \breve{a}-\sqrt{\kappa }\left( \breve{w}+\breve{b}_{in}\right).
\end{eqnarray*}%
Here, we have included the term  $\breve{w}$ which arises from indirect coupling to  an external system $W_f$ via the input field. For this system,  $\Omega_-=0$, $\Omega_+=\frac{i\epsilon}{2}$, $C_-=\sqrt{\kappa}$, and $C_+=0$. Now
$$
Q = \Delta( \kappa, -\epsilon) = \left[  \begin{array}{cc}
\kappa & -\epsilon
\\
-\epsilon & \kappa
\end{array} \right],
$$
and so $Q \geq 0$ if and only if $\epsilon \leq \kappa$. So the DPA is passive with performance variable $\breve{z} = -\sqrt{\kappa } \breve{a}$ if and only if $\epsilon \leq \kappa$, i.e., if and only if it is marginally or exponentially stable.

When $\epsilon < \kappa$, the $L^2$ gain from $\breve{w}$ to  $\breve{z}_{2} =\sqrt{\kappa }\breve{a}+\breve{w}$ is  $(\kappa + \epsilon)/(\kappa - \epsilon)$. This system is not lossless bounded real, but satisfies the physical realization criteria of section \ref{sec:models-realization}.

\subsubsection{Optical  Amplifier} \label{sec:analysis-eg-1-amp}

A model for an optical amplifier $G$ is  given  in \cite[page 215]{GZ04}, which corresponds to  a single open oscillator with $\Omega_-=0$, $\Omega_+=0$, $C_-=\sqrt{\kappa}$, $C_+=\sqrt{\gamma}$:
$$
\dot{\breve{a}} =-\frac{\kappa -\gamma }{2}\breve{a}-\left[
\begin{array}{cc}
\sqrt{\kappa } & -\sqrt{\gamma } \\
-\sqrt{\gamma } & \sqrt{\kappa }%
\end{array}%
\right] \left( \breve{w}+\breve{b}_{in}\right),
$$
where we have included the term  $\breve{w}$ which arises from indirect coupling to  an external system $W_f$ via the input field. For this model we have $Q=( \kappa - \gamma) I$, and so $G$ is exponentially stable if $\gamma < \kappa$, and marginally stable if $\gamma = \kappa$. Consequently, $G$ is passive with performance variable
$$\breve z =-\left[
\begin{array}{cc}
\sqrt{\kappa } & -\sqrt{\gamma } \\
-\sqrt{\gamma } & \sqrt{\kappa }%
\end{array}%
\right] \breve{a}
$$ if and only if $\gamma \leq \kappa $.

Finally, when the system $G$ is exponentially stable, the  $L^2$ gain of the transfer $\breve{w} \to \breve{z}_2 = \sqrt{\kappa} \, \breve{a} + \breve{w}$ is  $(\kappa + \gamma +2\sqrt{\kappa \gamma})/(\kappa - \gamma)$.

\subsubsection{Effect of Direct Coupling on $H^{\infty}$ Performance}
\label{sec:analysis-h-infty}

Consider two quantum systems $G_1$ and $G_2$  given by
\begin{eqnarray*}
\dot{a}_{1}(t) &=&-8a_{1}(t)-4w_1 (t) -4b_{in,1}(t), \\
b_{out,1}(t) &=&4a_{1}(t)+w_1 (t) +b_{in,1}(t), \\
\dot{a}_{2}(t) &=&-2a_{2}(t)-2w_2 (t) -2b_{in,2}(t), \\
b_{out,2}(t) &=&2a_{2}(t)+w_2 (t) +b_{in,2}(t).
\end{eqnarray*}%
Clearly, in both systems $C_{+}=\Omega _{+}=\Omega _{-}=0$. We suppose that these systems are directly coupled via a coupling of the form   (\ref{eq:direct-1}), with $K_-$ and $K_+$ being real.

\begin{figure}[tbh]
\epsfxsize=3.5in
\par
\epsfclipon
\par
\centerline{\epsffile{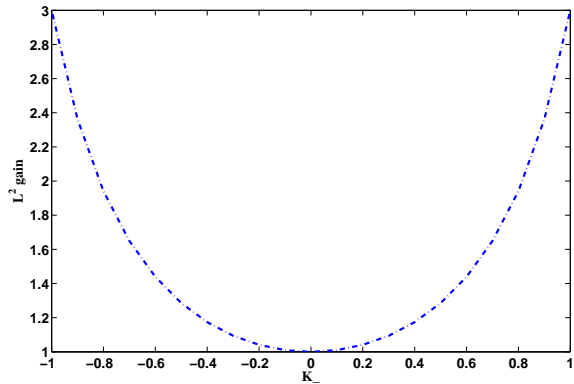}}
\caption{$L^2$ gain of the transfer $\breve{w}_1 \to \breve{z}_1$ in the example of section \ref{sec:analysis-h-infty} as a function of $K_{-}$ for fixed ratio $K_{+}= 3K_{-}$.}
\label{fig_Hinfty}
\end{figure}

If $K_{+} = 0$, the $\Omega_+$ term of the composite system is zero (c.f. (\ref{H})). Consequently,  by the discussion in section \ref{sec:analysis-brl}, the $H^{\infty }$ norm from $\breve{w}_{1}$ to $\breve{z}_1 = 4 \breve{a}_1 + \breve{w}_1$ is $1$. Next, fix the ratio $K_{+}/K_{-}$ to be $3$. Then the $L^2$ gain from $\breve{w}_{1}$ to $\breve{z}_1$ is plotted as a function of $K_{-}$ in Fig.~\ref{fig_Hinfty}. Clearly, this type of direct coupling
indeed influences the $L^2$ gain of the first channel of the system $G$.

\subsubsection{Effect of Direct Coupling on  LQG  Performance} \label{sec:analysis-LQG}

\begin{figure}[tbh]
\epsfxsize=3.5in
\par
\epsfclipon
\par
\centerline{\epsffile{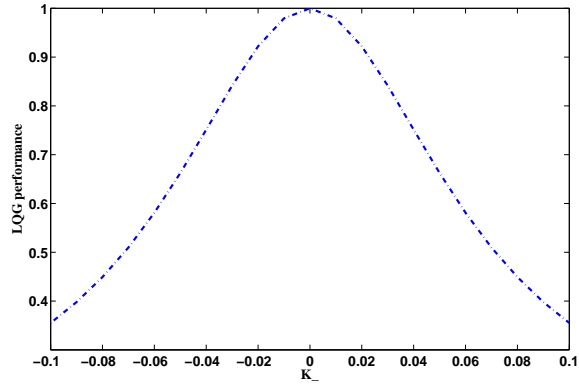}}
\caption{LQG performance for the example of section \ref{sec:analysis-LQG} as a function of $K_-$ while $K_{+} =0$.}
\label{fig_H2_a}
\end{figure}

\begin{figure}[tbh]
\epsfxsize=3.5in
\par
\epsfclipon
\par
\centerline{\epsffile{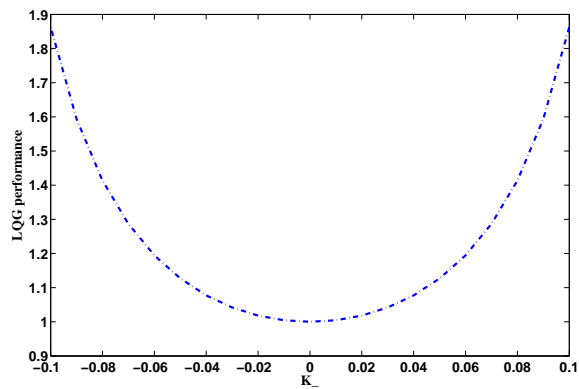}}
\caption{LQG performance for the example of section \ref{sec:analysis-LQG} as a function of $K_-$ with fixed ratio $K_{+}=1.1K_{-}$.}
\label{fig_H2_b}
\end{figure}

Given a linear  quantum system%
$$
\dot{a}_{1}(t) =-0.0187a_{1}(t) -0.1934b_{in,1}(t),
$$
a controller
$$
\dot{a}_{2}(t) =-0.2327a_{2}(t) -0.6822b_{in,2}(t),
$$
and a performance variable
$$
\breve{z}_{1}(t) =\breve{a}_{1}(t),
$$
in this section we show how direct coupling of the form (\ref{eq:direct-1}) can influence the infinite-horizon LQG cost defined as $\mathfrak{J}_{\infty}$ in (\ref{LQG_index}). For simplicity of exposition, assume both $K_-$ and $K_+$ are real numbers. Fix $K_{+} =0$, the LQG cost $\mathfrak{J}_{\infty}$ is plotted as a function of $K_- $ in Fig.~\ref{fig_H2_a}. Note that even when $K_{+} =0$, $\mathfrak{J}_{\infty}$ is affected by direct coupling.  Fix $K_{+}/K_{-} =1.1$, $\mathfrak{J}_{\infty}$ is plotted as a function of $K_- $ in Fig.~\ref{fig_H2_b}. These two examples clearly show that direct coupling is able to influence LQG performance specified for the plant.

\section{Coherent Feedback Controller Synthesis} \label{sec:synth}

So far we have looked at some basic performance characteristics of linear quantum systems (stability, passivity, gain), and in particular we have seen how direct coupling can influence  behavior. In this section we turn to the problem  of including  direct couplings  in systematic controller design methodologies.  Using direct couplings in design is natural from the physical point of view, and has been considered in \cite{Lloyd00}. Our interest here to design direct and indirect couplings to optimize  specific performance criteria. In general, explicit solutions are not known, and optimization algorithms are used.

We begin in section \ref{sec:closed_loop} with a description of the plant-controller feedback system to be used in the sequel. In section \ref{sec:synth-stab} we give a brief example to illustrate how direct coupling may be used to stabilize an otherwise unstable closed-loop system.   Sections \ref{sec:h-infty-synthesis} and \ref{sec:LQG-synthesis} extend the $H^\infty$ and LQG coherent controller synthesis methods of \cite{JNP08} and \cite{NJP09} respectively to include direct couplings.

\subsection{Closed-Loop Plant-Controller System}\label{sec:closed_loop}

\begin{figure}[htb]
\epsfxsize=3in
\par
\epsfclipon
\par
\centerline{\epsffile{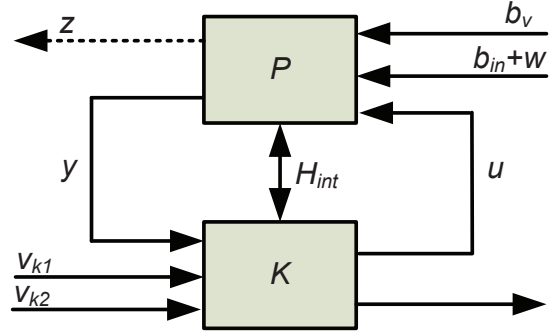}}
\caption{Coherent feedback control arrangement.}
\label{closed-loop}
\end{figure}

Consider two  quantum linear systems:   $P$, the plant to be controlled, coupled to  $K$, the controller, as shown in Figure \ref{closed-loop}. This feedback system involves both direct and indirect coupling between the plant and the controller. While the feedback architecture will be fixed, the parameters defining the controller (which includes the couplings) will be synthesized using $H^\infty$ and LQG performance criteria.

The plant $P$  is given by differential equations
\begin{eqnarray}
\dot{\breve{a}}(t) &=&A\breve{a}(t)+B_{12} \breve{a}_K(t)+
B_{v}\breve{b}_{v}(t)+B_{f}\breve{w}(t) \nonumber  \\
&&+B_{f}\breve{b}_{in}(t)+B_{u}\breve{u}(t), \, \, \breve{a}(0)=\breve{a}_{0},  \nonumber \\
\breve{y}(t) &=&C\breve{a}(t)+D_{v}\breve{b}_{v}(t)+D_{f}\breve{w}%
(t)+D_{f}\breve{b}_{in}(t).  \label{plant_optical}
\end{eqnarray}%
The inputs $\breve{w}(t)$ and $\breve{b}_{in}(t)$ are the same as defined in section \ref{sec:models-indirect}. $\breve{y}(t)$ is a selection of output field channels from the plant. $\breve{b}_{v}(t)$ is a vector of additional quantum noises; $\breve{u}(t)$ is a quantum field signal from the to-be-designed controller $K$, hence it is a vector of physical variables.  The term $B_{12} \breve{a}_K(t)$ are due to direct coupling which will be explained shortly.

A fully quantum controller used in this paper is a linear quantum system of the form\footnote{We assume that all the variables and matrices of the plant and the controller have compatible dimension, but we don't bother to specify them explicitly.}
\begin{eqnarray}
\dot{\breve{a}}_{K}(t) &=&A_{K}\breve{a}_{K}(t)+B_{21}\breve{a}(t)+B_{K}\breve{y}(t)+B_{K1}\breve{b}%
_{v_{K1}}(t) \nonumber \\
&&+B_{K2}\breve{b}%
_{v_{K2}}(t), \, \, \breve{a}_{K}(0)=\breve{a}_{K0},   \nonumber \\
\breve{u}(t) &=&C_{K}\breve{a}_{K}(t)+B_{K0}\breve{b}_{v_{K1}}(t).   \label{controller_optical2}
\end{eqnarray}%
This structure allows for direct coupling and indirect coupling between the plant $P$ and the controller $K$. Here, $\breve{b}_{v_{K1}}(t)$ and $\breve{b}_{v_{K2}}(t)$ are independent standard quantum white noises, and $\breve{u}(t)$ is the field output of the controller  corresponding to $\breve{b}_{v_{K1}}(t)$. Hence, $B_{K0}$ is a square matrix; in fact, it is an identity matrix.  Finally the terms $B_{12} \breve{a}_K(t)$ and  $B_{21}\breve{a}(t)$ is due to the direct coupling between the plant and controller in terms of an interaction Hamiltonian
\begin{equation}
\label{syn:Hamiltonian}
H_{PK}= \frac{1}{2}\left( \breve{a}^{\dagger }S^{\dagger }\breve{a}_{K}+%
\breve{a}_{K}^{\dagger }S\breve{a}\right),
\end{equation}
where
$
S=\Delta (iK_{-},iK_{+})
$
for complex matrices $K_{-}$ and $K_{+}$ of suitable dimensions.

The overall plant-controller system, including direct and indirect couplings, is
\begin{eqnarray}
\left[
\begin{array}{c}
\dot{\breve{a}}(t) \\
\dot{\breve{a}}_{K}(t)%
\end{array}%
\right]  & \hspace{-2mm} =&\hspace{-2mm} \left[
\begin{array}{cc}
A & B_{u}C_{K}+B_{12} \\
B_{K}C+B_{21} & A_{K}%
\end{array}%
\right] \left[
\begin{array}{c}
\breve{a}(t) \\
\breve{a}_{K}(t)%
\end{array}%
\right]   \nonumber \\
&&\hspace{-2mm} +\left[
\begin{array}{c}
B_{f} \\
B_{K}D_{f}%
\end{array}%
\right] \breve{w}(t)+ G_{cl} \left[
\begin{array}{c}
\breve{b}_{in}(t) \\
\breve{b}_{v}(t) \\
\breve{b}_{v_{K1}}(t)\\
\breve{b}_{v_{K2}}(t)%
\end{array}%
\right] ,  
  \label{syn_cls}
\end{eqnarray}
where $B_{12}$ and $B_{21}$ are given by (\ref{eq:direct-5}) and (\ref{eq:direct-7}) respectively and
$$
G_{cl}=\left[
\begin{array}{cccc}
B_{f} & B_{v} & B_{u}B_{K0} & 0\\
B_{K}D_{f} & B_{K}D_{v} & B_{K1} &B_{K2}%
\end{array}%
\right].
$$

The controller matrices $K_-, K_+$,  (or $B_{12}, B_{21}$), $A_K, B_K, C_K, B_{K1}, B_{K2}, B_{K0}$ are to be found to optimize performance criteria defined in terms of the  closed-loop performance variable
\begin{equation}
\breve{z}(t) =[ C_{p} ~~ D_{u}C_K ]\left[
\begin{array}{c}
\breve{a}(t) \\
\breve{a}_{K}(t) \\
\end{array}%
\right] +  \breve{D}_{pf}\breve{w}(t) .
\label{syn_performance}
\end{equation}%

Because standard matrix algorithms will be used in what follows, we may represent the plant-controller system in quadrature form as discussed in  section \ref{sec:models-quadrature}. Let $\tilde{a}$, $\tilde{a}_{K}$, $\tilde{w}$, $\tilde{b}_{in}$, $\tilde{b}_{v}$, $\tilde{u}$, $\tilde{z}$, $\tilde{y}$, $\tilde{b}_{v_{K1}}$,  $\tilde{b}_{v_{K2}}$  be the quadrature counterparts of $\breve{a}$, $\breve{a}_{K}$, $\breve{w}$, $\breve{b}_{in}$, $\breve{b}_{v}$, $\breve{z}$, $\breve{\beta}_{u}$, $\breve{y}$, $\breve{b}_{v_{K1}}$, $\breve{b}_{v_{K2}}$ respectively. Define
$$
\tilde{A}_{cl} =\left[
\begin{array}{cc}
\tilde{A} & \tilde{B}_{u}\tilde{C}_{K} \\
\tilde{B}_{K}\tilde{C} & \tilde{A}_{K}%
\end{array}%
\right]+\tilde{\Xi} , ~~\ \tilde{B}_{cl} = \left[
\begin{array}{c}
\tilde{B}_{f} \\
\tilde{B}_{K}\tilde{D}_{f}%
\end{array}%
\right] ,
$$
$$ \tilde{G}_{cl}=\left[
\begin{array}{cccc}
\tilde{B}_{f} & \tilde{B}_{v} & \tilde{B}_{u}\tilde{B}_{K0} & 0 \\
\tilde{B}_{K}\tilde{D}_{f} & \tilde{B}_{K}\tilde{D}_{v} & \tilde{%
B}_{K1} & \tilde{B}_{K2}%
\end{array}%
\right] ,
$$
$$
\tilde{C}_{cl} =\left[
\begin{array}{cc}
\tilde{C}_{p} & \tilde{D}_{u}\tilde{C}_{K}%
\end{array}%
\right] ,~~\tilde{D}_{cl} = \tilde{D}_{pf},
$$
where $\tilde{\Xi} =[0 ~ \tilde{B}_{12}; \tilde{B}_{21} ~ 0]$ satisfies $\tilde{B}_{21} = \Theta\tilde{B}_{12}^{T}\Theta$. Then the closed-loop system in the quadrature representation is given by%
$$
\left[
\begin{array}{c}
\dot{\tilde{a}}(t) \\
\dot{\tilde{a}}_{K}(t)%
\end{array}%
\right]  =\tilde{A}_{cl}\left[
\begin{array}{c}
\tilde{a}(t) \\
\tilde{a}_{K}(t)%
\end{array}%
\right] +\tilde{B}_{cl}\tilde{w}(t)+\tilde{G}_{cl}\left[
\begin{array}{c}
\tilde{b}_{in}(t) \\
\tilde{b}_{v}(t) \\
\tilde{b}_{v_{K1}}(t) \\
\tilde{b}_{v_{K2}}(t)%
\end{array}%
\right] ,
$$
$$
\tilde{z}(t) =\tilde{C}_{cl}\left[
\begin{array}{c}
\tilde{a}(t) \\
\tilde{a}_{K}(t)%
\end{array}%
\right]+\tilde{D}_{cl}\tilde{w}(t) .
$$

\begin{remark}\label{rm:B_K0} {\rm
Since $B_{K0}$ is an identity matrix, $\tilde{B}_{K0} = I$. Note that the coefficient matrix of $d v_K$ in \cite[Eq. (23)]{JNP08} is in fact $[I ~ 0]$ because $v_K = [v_{K1}^T ~ v_{K2}^T]^T$ is used there.
}
$\Box$
\end{remark}

\subsection{Stabilization} \label{sec:synth-stab}

The following example shows that an appropriate direct coupling is sufficient to stabilize an unstable closed-loop system formed by indirect coupling via field channels. The example is constructed in the following way: Let $\Omega_- = 0$ and $\Omega_+ =0$ for both the plant and the controller. Then for fixed $C_-$, choose $C_+$ for both the plant and the controller such that each of them is stable while the closed-loop system is unstable. Finally find parameters $K_-$ and  $K_+$ of direct coupling such that the closed-loop system is stable. The example is of illustrative nature, with no practical intention.


Consider a  quantum linear feedback system as  in Fig.~\ref{closed-loop} where the quantum plant is  given by
\begin{eqnarray}
\dot{\breve{a}}(t) &=& -0.1331 \breve{a}(t) +\left[
\begin{array}{cc}
-0.3420 & 0.2897 \\
0.2897 & -0.3420%
\end{array}%
\right] \breve{b}_{in}(t) \nonumber \\
&& +\left[
\begin{array}{cc}
-0.8180 & 0.6602 \\
0.6602 & -0.8180%
\end{array}%
\right] \breve{u}(t),  \nonumber \\
\breve{y}(t) &=&\left[
\begin{array}{cc}
0.3420 & 0.2897 \\
0.2897 & 0.3420%
\end{array}%
\right] \breve{a}(t)+\breve{b}_{in}(t).  \label{Ex1}
\end{eqnarray}%
Here,  $\breve{y}(t)$ is the
field output corresponding to  $\breve{b}_{in}(t)$.
The indirect coupling is given by
\begin{eqnarray}
\dot{\breve{a}}_{K}(t) &=&-0.1321 \breve{a}_{K}(t)+\left[
\begin{array}{cc}
-0.7271 & 0.3093 \\
0.3093 & -0.7271%
\end{array}%
\right] \breve{y}(t) \nonumber \\
&&+\left[
\begin{array}{cc}
-0.3412 & 0.5341 \\
0.5341 & -0.3412%
\end{array}%
\right] \breve{v}_{K1}(t),  \nonumber \\
\breve{u}(t) &=&\left[
\begin{array}{cc}
0.3412 & 0.5341 \\
0.5341 & 0.3412%
\end{array}%
\right] \breve{a}_{K}(t)+\breve{v}_{K1}(t).  \label{Ex1-2}
\end{eqnarray}%
That is, $\breve{u}(t)$ is the field output of $K$ corresponding to $\breve{v}_{K1}(t)$. It can be shown the closed-loop $A$-matrix is not Hurwitz despite the fact that both of the $A$-matrices of the plant and the controller are
indeed Hurwitz.

Now if the direct coupling is  specified by $K_{-} = -0.6$ and $K_{+} = - 0.42$, it can be verified that the new closed-loop $A$-matrix is Hurwitz.

\subsection{$H^{\infty}$ Synthesis} \label{sec:h-infty-synthesis}

In section \ref{sec:analysis-h-infty} a simple example shows that direct coupling can influence the $L^2$ gain of channels in linear quantum systems, and the purpose of this section is to include this additional design flexibility into the $H^\infty$ synthesis methodology of \cite{JNP08}. Before going into the more general set-up, let us first illustrate how to design direct and indirect couplings to affect the input-output behavior of a given quantum plant.

\subsubsection{A Simple Example for $H^{\infty}$ Synthesis} \label{syn:simple_example}

Consider a quantum plant of the form%
\begin{eqnarray*}
\dot{a}(t) &=&-\frac{\gamma }{2}a(t)-\sqrt{\kappa _{1}}\left(
w_{1}(t)+b_{in,1}(t)\right) -\sqrt{\kappa _{2}}b_{in,2}(t), \\
z(t) &=&\sqrt{\kappa _{2}}a(t), \,\,a(0)=a,\,\,%
\gamma =\kappa _{1}+\kappa _{2}.
\end{eqnarray*}%
Here, the disturbance $w_1$ is an external variable entering the first field channel, while the performance variable $z$ is a component of the output of the second field channel $b_{out,2}(t) = z(t) + b_{in,2}(t)$.

We assume the indirect coupling is an open optical cavity interacting with one field channel%
\[
\dot{a}_{K}(t)=\left( -i\omega -\frac{\kappa_{3}}{2}\right) a_{K}(t)-\sqrt{\kappa_{3}}%
b_{v_{K1}}(t),\,\,a_{K}(0)=a_{K},
\]%
where $\omega $ and $\kappa_{3}$ are real. Adding a particular direct coupling of the form (\ref{eq:direct-1}) where  $S=\Delta \left( iK_{-},0\right) $ with $K_{-}$ being real for simplicity, then the closed-loop system is (in the quadrature representation)
\begin{eqnarray*}
\dot{\tilde{a}}(t) &=&-\frac{\gamma }{2}\tilde{a}(t)-K_{-}\tilde{a}_{K}(t)\\
&&-\sqrt{\kappa _{1}}\left( \tilde{w}_{1}(t)+\tilde{b}_{in,1}(t)\right) -\sqrt{%
\kappa _{2}}\tilde{b}_{in,2}(t), \\
\dot{\tilde{a}}_{K}(t) &=&\left( \omega \Theta_1 -\frac{\kappa _{3}}{2}\right) \tilde{a}%
_{K}(t)+K_{-}\tilde{a}(t) -\sqrt{\kappa_{3}}
\tilde{b}_{v_{K1}}(t), \\
\tilde{z}(t) &=&\sqrt{\kappa _{2}}\tilde{a}(t)+\tilde{b}_{in,2}(t),
\end{eqnarray*}%
where $\Theta_1 =[0 ~ 1; -1 ~ 0]$.
The closed loop system matrices are
\[
\tilde{A}_{cl}=\left[
\begin{array}{cc}
-\frac{\gamma }{2}I & -K_{-}I \\
K_{-}I & \omega \Theta_1 -\frac{\kappa_{3}}{2}I%
\end{array}%
\right] , ~ \tilde{B}_{cl}=\left[
\begin{array}{c}
-\sqrt{\kappa _{1}}I \\
0%
\end{array}%
\right] ,
\]
\[
\tilde{C}_{cl}=\left[
\begin{array}{cc}
\sqrt{\kappa _{2}}I & 0%
\end{array}%
\right] , ~ \tilde{D}_{cl}=0 .
\]%

We now seek to minimize the $L^2$ gain of the transfer $\breve{w}_1 \to \breve{z}$ with respect to the direct coupling parameter $K_-$. Fix $\kappa _{1}=2$, $\kappa _{2}=5$, and $\kappa_{3} = 3$. Without direct coupling, the $L^2$ gain is $0.9035$.  Note that the unknown variables are real numbers $\omega $ (for indirect coupling) and $K_{-}$ (for direct coupling). Matlab function `fminsearch' yields a local minimal value of  the $L^2$ gain $0.6284$ at $\omega =0
$ and $K_{-}=17.7135$.

\begin{remark}{\rm
If $\kappa_{3}$ goes to $\infty$, it can be shown that  the $L^2$ gain  approaches $0$. The physical reason for this is that when $\kappa_3$ is large, the coupling of the controller to the controller field channel is strong, and energy entering the first field channel exists via the controller output channel.
} $\Box$
\end{remark}

\subsubsection{LMI Formulation}

The preceding section shows that certain simple optimization functions can be used to design direct couplings to alter the input-output behavior of quantum plants.  In the section we present a general formulation using LMIs.

According to the strict bounded real lemma (Theorem \ref{thm:brl}), the closed-loop is internally stable and strictly bounded real with disturbance attenuation $g$ if and only if there is a real symmetric matrix $\mathcal{P}$ such that%
\begin{eqnarray}
\mathcal{P} & > & 0  \label{LMI} \\
\left[
\begin{array}{ccc}
\tilde{A}_{cl}^{T}\mathcal{P}+\mathcal{P}\tilde{A}_{cl} & \mathcal{P}\tilde{B}_{cl} &
\tilde{C}_{cl}^{T} \\
\tilde{B}_{cl}^{T}\mathcal{P} & -gI & \tilde{D}_{cl}^{T} \\
\tilde{C}_{cl} & \tilde{D}_{cl} & -gI%
\end{array}%
\right] &<&0.  \label{LMI2}
\end{eqnarray}%
In what follows controller parameters $\tilde{A}_{K}$, $\tilde{B}_{K}$, $\tilde{C}_{K}$ and direct coupling parameters $\tilde{\Xi} $ are derived based on a multi-step optimization procedure. Observe that inequality (\ref{LMI2}) is nonlinear, a change of variables technique is proposed in the literature to convert it to linear matrix inequalities \cite{SGCr97}. We outline this technique briefly. Following the development in \cite{SGCr97}, decompose $\mathcal{P}$ and its inverse $\mathcal{P}^{-1}$ as
$$
\mathcal{P=}\left[
\begin{array}{cc}
\mathbf{Y} & N \\
N^{T} & \ast%
\end{array}%
\right] , ~~ \mathcal{P}^{-1}=\left[
\begin{array}{cc}
\mathbf{X} & M \\
M^{T} & \ast%
\end{array}%
\right] .
$$%
Define%
$$
\Pi _{1}=\left[
\begin{array}{cc}
\mathbf{X} & I \\
M^{T} & 0%
\end{array}%
\right] , ~~ \Pi _{2}=\left[
\begin{array}{cc}
I & \mathbf{Y} \\
0 & N^{T}%
\end{array}%
\right] .
$$%
And also define a change of variables%
\begin{eqnarray}
\mathbf{\hat{A}} &\mathbf{=}&N( \tilde{A}_{K}M^{T}+\tilde{B}_{K}%
\tilde{C}\mathbf{X}) +\mathbf{Y}( \tilde{B}_{u}\tilde{C}%
_{K}M^{T}+\tilde{A}\mathbf{X}) ,  \nonumber \\
\mathbf{\hat{B}} &\mathbf{=}&N\tilde{B}_{K},  \nonumber \\
\mathbf{\hat{C}} &\mathbf{=}&\tilde{C}_{K}M^{T},  \nonumber \\
\mathbf{\Omega } &=&\Pi _{1}^{T}\mathcal{P}\tilde{\Xi} \Pi _{1}.   \label{congruence}
\end{eqnarray}%
With these notations, (\ref{LMI2}) holds if and only if the following inequalities holds.%
\begin{equation}
-\left[
\begin{array}{cc}
X & I \\
I & Y%
\end{array}%
\right] <0 ,  \label{LMIs2}
\end{equation}%
\begin{eqnarray}
&& \hspace{-6mm} \left[
\begin{array}{c}
\tilde{A}\mathbf{X+X}\tilde{A}^{T}+\tilde{B}_{u}\mathbf{\hat{C}+}(\tilde{B}%
_{u}\mathbf{\hat{C}})^{T} \\
\mathbf{\hat{A}+}\tilde{A}^{T} \\
\tilde{B}_{f}^{T} \\
\tilde{C}_{p}\mathbf{X+}\tilde{D}_{u}\mathbf{\hat{C}}%
\end{array}%
\right.   \nonumber \\
&&\hspace{1cm} \left.
\begin{array}{ccc}
\tilde{A}+\mathbf{\hat{A}}^{T} & \ast  & \ast  \\
\tilde{A}^{T}\mathbf{Y}+\mathbf{Y}\tilde{A}+\mathbf{\hat{B}}\tilde{C}+(%
\mathbf{\hat{B}}\tilde{C})^{T} & \ast  & \ast  \\
(\mathbf{Y}\tilde{B}_{f}+\mathbf{\hat{B}}\tilde{D}_{f})^{T} & -gI & \ast  \\
\tilde{C}_{p} & \tilde{D}_{cl} & -gI%
\end{array}%
\right]  \nonumber \\
&&\hspace{-8mm}+\left[
\begin{array}{cc}
\tilde{B}_{12}M^{T}+(\tilde{B}_{12}M^{T})^{T} & (N\tilde{B}_{21}\mathbf{X}%
)^{T}+(\mathbf{Y}\tilde{B}_{12}M^{T})^{T} \\
N\tilde{B}_{21}\mathbf{X}+\mathbf{Y}\tilde{B}_{12}M^{T} & N\tilde{B}_{21}+(N%
\tilde{B}_{21})^{T} \\
0 & 0 \\
0 & 0%
\end{array}%
\right.  \nonumber  \\
&& \hspace{6.5cm} \left.
\begin{array}{cc}
0 & 0 \\
0 & 0 \\
0 & 0 \\
0 & 0%
\end{array}%
\right]  \nonumber  \\
&<&0. \label{LMIs}
\end{eqnarray}
If (\ref{LMIs2}) and (\ref{LMIs}) are simultaneously soluble, then a controller and the interaction Hamiltonian can be obtained. More specifically, according to Eq. (\ref{congruence}), the following matrices can be obtained:%
\begin{eqnarray}
\tilde{B}_{K} & \hspace{-2mm}  =& \hspace{-2mm} N^{-1}\mathbf{\hat{B}},  \label{solution} \\
\tilde{C}_{K} & \hspace{-2mm}  =& \hspace{-2mm} \mathbf{\hat{C}}\left( M^{T}\right) ^{-1},  \nonumber \\
\tilde{A}_{K} & \hspace{-2mm}  =& \hspace{-2mm} N^{-1}( \mathbf{\hat{A}-}N\tilde{B}_{K}\tilde{C}\mathbf{X-Y}( \tilde{B}_{u}\tilde{C}_{K}M^{T}+\tilde{A}\mathbf{X}))M^{-T},  \nonumber \\
\tilde{\Xi} & \hspace{-2mm}  =& \hspace{-2mm} \mathcal{P}^{-1}\left( \Pi _{1}^{-T}\right)\mathbf{\Omega }\Pi
_{1}^{-1}.  \label{interaction}
\end{eqnarray}%
Unfortunately, notice that there are such terms as $N\tilde{B}_{21}\mathbf{X}$ and $\mathbf{Y}\tilde{B}_{12}M^{T}$ in inequality (\ref{LMIs}), which make it still nonlinear. The above analysis shows it is hard to directly utilize LMI techniques to do controller design when direct coupling is involved.

\subsubsection{Multi-step Optimization} \label{algorithm}

In this section, we intend to circumvent the above difficulty based on a multi-step optimization procedure which is formulated as follows:

\textit{Initialization}. Set $\tilde{B}_{12 }=0$ and $\tilde{B}_{21 }=0$.

\textit{Step 1}. Employ LMI techniques to solve linear matrix inequalities (\ref{LMIs2}) and (\ref{LMIs}). Then choose matrices $M$ and $N$ satisfying $MN^{-1}=I-XY$. With these parameters, indirect coupling parameters parameters $\tilde{A}_{K}$, $\tilde{B}_{K}$, and $\tilde{C}_{K}$ are obtained via (\ref{solution}).

\textit{Step 2. } Pertaining to \textit{Step 1}. Solve inequalities in (\ref{LMIs}) to find direct coupling parameters $\tilde{B}_{12 }$ and disturbance gain $g$.

\textit{Step 3. } Fix $\tilde{B}_{12 }$ and $\tilde{B}_{21 }$ to those obtained in \textit{Step 2} and fix $M $ and $N$ obtained in \textit{Step 1}, restart from \textit{Step 1 }to find parameters $\mathbf{\hat{A}}$, $\mathbf{\hat{B}}$, $\mathbf{\hat{C}}$, $\mathbf{X}$, $\mathbf{Y}$, and disturbance gain $g$. Then go to \textit{Step 2}.

After this iterative procedure is complete, use the values  $B_{12}, B_{21}$, $A_K, B_K, C_K$ obtained to find $B_{K1}, B_{K2}, B_{K0}$ to ensure physical realizability of the controller (see section \ref{sec:physical-realization} for details).

\begin{remark}
{\rm \textit{Steps 1 and 2 } are standard LMI problems which can be solved efficiently using the Matlab LMI toolbox. However, there is some delicate issue in \textit{Step 3.} Assume that $\tilde{B}_{12 }$ and $\tilde{B}_{21 }$ have been obtained in \textit{Step 2}. According to the second item in (\ref{LMIs}), constant matrices $M$ and $N$ must be specified in order to render (\ref{LMIs}) linear in parameters $\mathbf{\hat{A}}$, $\mathbf{\hat{B}}$, $\mathbf{\hat{C}}$, $\mathbf{X}$, $\mathbf{Y}$, and disturbance gain $g$. In \textit{Step 3}, $M$ and $N$ obtained in \textit{Step 1 } is used. Unfortunately, this choice of $M$ and $N$ sometimes may generate a controller whose parameters are ill-conditioned. Due to this reason, $M$ and $N$ in \textit{Step 3} might have to be chosen carefully to produce a physically meaningful controller. This fact is illuminated by an example in Section \ref{ex:physi}.
}
\end{remark}

\subsubsection{Physical Realizability}
\label{sec:physical-realization}

Because the to-be-designed controller is fully quantum, its dynamical evolution must obey laws of quantum mechanics, as a result controller parameter matrices cannot be chosen arbitrarily. This is the so-called physical realizability (of fully quantum controllers) as discussed in section \ref{sec:models-realization}. In this section we investigate  this issue in more detail by focusing on the structure of the controller (\ref{controller_optical2}). Firstly, direct coupling demands equalities (\ref{eq:direct-5}) and (\ref{eq:direct-7}) for certain matrices $K_-$ and $K_+$ for an interaction Hamiltonian $H_{PK}$ defined in (\ref{syn:Hamiltonian}). Secondly, equations (\ref{eq:c-2-ccr})-(\ref{eq:c-2-f}) for indirect coupling can be
rewritten as%
\begin{eqnarray}
J_{n} A+A^{\dagger }J_{n}+\left( B_{f}^{\flat }\right) ^{\dagger}J_{m}B_{f}^{\flat } &=&0 , \label{eq:c-2-ccr2} \\
C_{f} &=&-B_{f}^{\flat } .  \label{eq:c-2-f-2}
\end{eqnarray}%
Since the input of the quantum controller (\ref{controller_optical2}) is partitioned into three blocks, it is better to rewrite the above relations as
\begin{eqnarray}
J_{n_{K}} A_{K}+A_{K}^{\dagger }J_{n_{K}} + ( B_{K1}^{\flat }) ^{\dagger } J_{m_{1}}  B_{K1}^{\flat }  & & \nonumber\\
+ ( B_{K2}^{\flat }) ^{\dagger } J_{m_{2}} B_{K2}^{\flat } + \left( B_{K}^{\flat }\right) ^{\dagger } J_{m_{3}} B_{K}^{\flat }&=& 0, \label{LQG:physi-reali} \\
C_{K} &=&-B_{K1}^{\flat }, \label{LQG:physi-reali_b} \\
B_{K0} &=&I.
\label{LQG:physi-reali_c}
\end{eqnarray}

Similarly, in the quadrature representation the controller is physically realizable if and only if
\begin{eqnarray}
\tilde{A}_{K}\Theta_{n_{K}}+\Theta_{n_{K}}\tilde{A}_{K}^{T}+ \tilde{B}_{K1} \Theta_{m_{1}}\tilde{B}_{K1}^{T}  & & \nonumber \\
+ \tilde{B}_{K2} \Theta_{m_{2}}\tilde{B}_{K2}^{T} + \tilde{B}_{K} \Theta_{m_{3}}\tilde{B}_{K}^{T} &=& 0, \label{LQG:physi-reali_real} \\
\tilde{C}_{K} &=&\Theta_{m_{1}} \tilde{B}_{K1}^{T}\Theta_{n_{K}}, \nonumber \\
\tilde{B}_{K0} &=&I  \nonumber \\
\tilde{B}_{21} &=& \Theta_{n_{K}}\tilde{B}_{12}^{T}\Theta_n.  \nonumber
\end{eqnarray}
hold, where the subscript $n$ is the dimension of $a(t)$, and $\tilde{B}_{12}$ is an arbitrary matrix.

The issue of physical realizability of indirect coupling has been addressed in \cite{JNP08} (see, e.g., \cite[Theorem 5.5 and Lemma 5.6]{JNP08}), where it is shown that, for arbitrarily given $\tilde{A}_{K}$, $\tilde{B}_{K}$, and $\tilde{C}_{K}$, there always exist $\tilde{B}_{K1}$, $\tilde{B}_{K2}$, and $\tilde{B}_{K0}$ such that the resulting indirect coupling is physically realizable. Noticing the fact that unitary transformation defined in section \ref{sec:models-quadrature} is equivalent to the transformation in \cite[section III. A.]{JNP08} in terms of similarity transformations, it is easy to show that controllers designed via the multi-step optimization procedure can be
indeed fully quantum. The procedure goes like this. Firstly, after \textit{step 1} is implemented, a similar technique like that in \cite[Lemma 5.6]{JNP08} can be used to obtain an indirect coupling which is physically realizable. Second,  after \textit{step 1} is implemented, $\tilde{B}_{12 }$ is obtained. Let $\tilde{B}_{21} =\Theta_{n_{K}}\tilde{B}_{12}^{T}\Theta_n$. Then  a direct coupling is constructed. Clearly, the quantum controller composed of the indirect coupling and direct coupling is physically realizable. Finally, since direct coupling does not affect indirect coupling,  \textit{Steps 3} can always yield a physically  realizable indirect coupling.  Consequently, the multi-step optimization procedure indeed is able to produce a fully quantum controller.

\subsubsection{Example 1}

In this section, an example is studied to demonstrate the effectiveness of the multi-step optimization approach proposed in the preceding section. Interestingly, due to the special structure of matrices, a physically realizable quantum controller can be constructed directly, without going through a procedure like that in \cite[Lemma 5.6]{JNP08}.

The following optical cavity system is studied in \cite{JNP08}.
\begin{eqnarray*}
\dot{a}(t) &=&-\frac{\gamma }{2}a(t)-\sqrt{\kappa _{1}}A_{1}(t)-\sqrt{\kappa
_{2}}A_{2}(t)-\sqrt{\kappa _{3}}A_{3}(t),  \\
\dot{a}^{\ast }(t) &=&-\frac{\gamma }{2}a^{\ast }(t)-\sqrt{\kappa _{1}} A_{1}^{\ast }(t)-\sqrt{\kappa _{2}}A_{2}^{\ast }(t)-\sqrt{\kappa _{3}} A_{3}^{\ast }(t), \\
\dot{B}_{3}(t) &=&\sqrt{\kappa _{3}}a(t)+A_{3}(t), \\
\dot{B}_{2}(t) &=&\sqrt{\kappa _{2}}a(t)+A_{2}(t), \, a(0)=a,  \, a^{\ast }(0)=a^{\ast },
\end{eqnarray*}%
where $\gamma =\kappa _{1}+\kappa _{2}+\kappa _{3}$. $\kappa _{1}$, $\kappa_{2}$, and $\kappa _{3}$ are coupling coefficients. In our notation,  $A_{1}(t)$, $A_{2}(t)$, $A_{3}(t)$, $B_{2}(t)$, and $B_{3}(t)$ correspond to $b_{v}(t)$, $w(t)+b_{in} (t)$, $u(t)$, $y(t)$, and $z(t)$ in the formulation outlined in section \ref{sec:closed_loop} respectively. The control problem studied in \cite{JNP08} is to minimize the influence of $A_{2}(t)$ on $B_{3}(t)$. In what following we investigate robustness of the control system.  Let $\kappa_{1}=2.6,\kappa _{2}=\kappa _{3}=0.2$. Suppose there is no direct coupling. Fix $g=0.1$. Then solving (\ref{LMIs2}) and (\ref{LMIs}) yields an $H^{\infty }$ suboptimal controller with parameters  $\tilde{A}_{K} = -3.0803 I$, $\tilde{B}_{K} =0.6801 I$, and $\tilde{C}_{K}= 0.5180 I$. This controller provides a disturbance attenuation level $0.0487$. Now suppose the optical cavity parameter $\kappa _{1}$ suffers from uncertainty \cite{BMR09}; for example, its effective value is $1.3$, instead of $2.6$, then the original controller yields a disturbance attenuation level $0.1702$, a significant performance degradation. Adding a direct coupling can improve this situation. In fact, by implementing \textit{step 2 }in the multi-step optimization approach proposed above, direct coupling matrices
$$
\tilde{B}_{12 }=\left[
\begin{array}{cc}
0.1648 &  -4.1842 \\
4.1842 &  0.1648%
\end{array}%
\right],\tilde{B}_{21 }=\left[
\begin{array}{cc}
-0.1648 &  -4.1842 \\
4.1842  &  -0.1648%
\end{array}%
\right]
$$%
are obtained. With such direct coupling the resulting disturbance attenuation level is $0.0618$, which is close to the original $0.0487$, a significant improvement over that involving indirect coupling solely. Finally, assume $y$ first passes through a $180^{0}$ phase shifter $(e^{i\pi})$  before it is sent to the controller, then the physically realizable controller (in the annihilation-creation representation) is given by
\begin{eqnarray*}
\dot{\breve{a}}_{K}(t) &=&-3.0803\breve{a}_{K}(t)-0.1648\breve{a}(t)  \\
&& +4.1842i\left[
\begin{array}{cc}
1 & 0 \\
0  &  -1
\end{array}%
\right]\breve{a}(t) -0.6801\breve{y}(t) \\
&&-0.5180\breve{b}%
_{v_{K1}}(t)- 2.3302\breve{b}%
_{v_{K2}}(t),    \\
\breve{u}(t) &=&0.5180\breve{a}_{K}(t)+\breve{b}_{v_{K1}}(t),~~ \breve{a}_{K}(0)=\breve{a}_{K0}.
\label{controller_optical}
\end{eqnarray*}%

\subsubsection{Example 2} \label{ex:physi}


In what follows the example discussed in \cite{NJP09} is re-studied to demonstrate how to use a technique similar to that in \cite[Lemma 5.6]{JNP08} to yield a physically realizable quantum controller.

Consider a linear quantum plant described by a set of quantum stochastic differential equations\footnote{In this example and all the examples below, the quantum plant is a single quantum oscillator. Consequently, $\tilde{a}$ is the same as $x$ in \cite{JNP08}, and all systems matrices are the same as those defined in \cite{JNP08}.}:
\begin{eqnarray}
\dot{\tilde{a}}(t) &=&\left[
\begin{array}{cc}
0 & \Delta \\
-\Delta & 0%
\end{array}%
\right] \tilde{a}(t)+\left[
\begin{array}{cc}
0 & 0 \\
0 & -2\sqrt{k_{1}}%
\end{array}%
\right] \tilde{u}(t)\nonumber\\
&&+\left[
\begin{array}{cc}
0 & 0  \\
0 & -2\sqrt{k_{2}}%
\end{array}%
\right](\tilde{b}_{in,1}(t)+\tilde{w}_{1}(t) )\nonumber\\
&&+\left[
\begin{array}{cc}
0 & 0 \\
 0 & -2\sqrt{k_{3}}%
\end{array}%
\right]\tilde{b}_{in,2 }(t) ,  \nonumber \\
\tilde{y}(t) &=&\left[
\begin{array}{cc}
2\sqrt{k_{2}} & 0 \\
0 & 0%
\end{array}%
\right] \tilde{a}(t)+\left[
\begin{array}{cc}
1 & 0 \\
0 & 1 %
\end{array}%
\right]\tilde{b}_{1in }, \label{example}
\end{eqnarray}%
where $\Delta =0.4$, $k_{1}=k_{2}=0.2$, and $k_{3}=0.4$.  Let the performance variable be that defined above with $\tilde{C}_{p}$ and $\tilde{D}_{u}$ being identity and  $\tilde{D}_{pf}$ being zero.

Assume $w_{1}$ is a light shone in the cavity. In what follows a controller is to be designed to minimize the $H^{\infty }$ norm from $\tilde{w}_{1}$ to $\tilde{z}$. First of all, implementing step 1 yields an indirect coupling with parameters
$$
\tilde{A}_{K}=\left[
\begin{array}{cc}
0.7372 & 1.3695 \\
-6.5299 & -7.0978%
\end{array}%
\right],
$$
$$
\tilde{B}_{K}=\left[
\begin{array}{cc}
-2.5963 & -1.3831 \\
-0.0012 & 9.7887%
\end{array}%
\right] ,
$$
$$
\tilde{C}_{K}=\left[
\begin{array}{cc}
0.1366 & 0.0423 \\
-0.6401 & -0.7339%
\end{array}%
\right] .
$$%
With this controller (more precisely, indirect coupling), the resulting $H^{\infty }$ norm is $1.7252$. Secondly, step 2 is used to obtain a direct coupling parameterized by
$$
\tilde{B}_{12 }=\left[
\begin{array}{cc}
0.0336  &  0.0128 \\
0.0403  &  0.0403%
\end{array}\right], \tilde{B}_{21 }=\left[
\begin{array}{cc}
-0.0403  &  0.0128 \\
0.0403   &  -0.0336%
\end{array}%
\right],
$$%
which yields an $H^{\infty }$ norm with value $1.6889$. Finally, fix $M$ and $N$ obtained in Step 1, Solving (\ref{LMIs2}) and (\ref{LMIs}) produces another indirect coupling with parameters
$$
\tilde{A}_{K}=\left[
\begin{array}{cc}
-0.0001 & 0.4002 \\
-0.1604 & -3.6773%
\end{array}%
\right] ,
$$
$$
\tilde{B}_{K}=10^{4}\times \left[
\begin{array}{cc}
-0.0001 & -0.0001 \\
-0.0000 & 1.2458%
\end{array}%
\right] ,
$$
$$
\tilde{C}_{K}=10^{-3}\times \left[
\begin{array}{cc}
0.0718 & 0.0000 \\
0.0192 & -0.2952%
\end{array}%
\right] .
$$%
Clearly, this controller is ill-conditioned. On the other hand, let $M=N$ be identity matrix. Solving (\ref{LMIs2}) and (\ref{LMIs}) produces another indirect coupling with parameters
$$
\tilde{A}_{K}=\left[
\begin{array}{cc}
0.0863 & 0.4907 \\
-20.1977 & -20.2797%
\end{array}%
\right],
$$
$$\tilde{B}_{K}=\left[
\begin{array}{cc}
-1.3563 & -0.6781 \\
-0.0003 & 162.3435%
\end{array}%
\right],
$$
$$\tilde{C}_{K}=\left[
\begin{array}{cc}
0.0062 & 0.0005 \\
-0.1226 & -0.1255%
\end{array}%
\right] .
$$%
With this controller (with both direct and indirect couplings), the resulting $H^{\infty }$ norm is $1.6056$. If $[
\begin{array}{cc}
\tilde{B}_{K} & \tilde{B}_{21}%
\end{array}%
] $ and $[
\begin{array}{cc}
\tilde{y}^{T}(t) & \tilde{a}^{T}(t)%
\end{array}%
] ^{T}$ are identified with $\tilde{B}_{K}$ and $y(t)$ in \cite[Lemma 5.6]{JNP08} respectively, then following \cite[Lemma 5.6]{JNP08} one has
$$
\tilde{B}_{K1}=\left[
\begin{array}{cc}
0.1255 & 0.0005 \\
-0.1226 & -0.0062%
\end{array}%
\right] , ~ \tilde{B}_{K2} = -15.5039 I.
$$%
 The resulting fully quantum controller is
\begin{eqnarray*}
\dot{\tilde{a}}_{K} (t) &=&\left[
\begin{array}{cc}
0.0863 & 0.4907 \\
-20.1977 & -20.2797%
\end{array}%
\right] \tilde{a}_{K} (t)\\
&&+\left[
\begin{array}{cc}
-0.0403 & 0.0128 \\
0.0403 & -0.0336%
\end{array}%
\right] \tilde{a}(t)\\
&&+\left[
\begin{array}{cc}
0.1255 & 0.0005 \\
-0.1226 & -0.0062%
\end{array}%
\right] \tilde{v}_{K1}(t) \\
&&+\left[
\begin{array}{cc}
-15.5039 & 0 \\
0 & -15.5039%
\end{array}%
\right] \tilde{v}_{K2}(t)\\
&&+\left[
\begin{array}{cc}
-1.3563 & -0.6781 \\
-0.0003 & 162.3435%
\end{array}%
\right] \tilde{y}(t); \\
\tilde{u}(t) &=&\left[
\begin{array}{cc}
0.0062 & 0.0005 \\
-0.1226 & -0.1255%
\end{array}%
\right] \tilde{a}_{K} (t)+\tilde{v}_{K1}(t),
\end{eqnarray*}%
with $ \tilde{a}_{K} (0) = \tilde{a}_{K}$. Clearly, $\tilde{u}(t)$ is the field output corresponding to field input $\tilde{v}_{K1}(t)$ (in vacuum state), while $\tilde{v}_{K2}(t)$ is an extra field input in vacuum state.

\begin{remark}{\rm
On the one hand, this example illustrates the effectiveness of the proposed multi-step optimization procedure; on the other hand, it reveals some delicate issue in this procedure like the choice of $M$ and $N$; some caution must be taken. }
$\Box$
\end{remark}

\subsection{LQG Synthesis} \label{sec:LQG-synthesis}

In this section we study the problem of coherent quantum LQG synthesis by means of both direct and indirect couplings. A simple example is first discussed which shows that, for simple quantum plants, general-purpose optimization methods can be used to tackle this problem. However, for more complex quantum plants, more systematic methods have to be developed.

\subsubsection{A Simple Example for LQG Synthesis}\label{syn:LQG_simple_example}

Given an open quantum plant parameterized by $\Omega _{-}=1, \, \, \Omega _{+}=2, \, \, C_{-}=\sqrt{2}, \, \, C_{+}=0$, and a closed quantum controller parameterized by $\Omega _{-}^{(2)}=2, \, \, \Omega _{+}^{(2)}=\sqrt{2}, \, \, C_{-}^{(2)}=0, \, \, C_{+}^{(2)}=0$, assume a direct coupling of the form (\ref{eq:direct-1}) where $K_{-}$ and $K_{+}$ are real numbers for simplicity. The resulting closed-loop system is
\begin{eqnarray*}
\dot{\breve{a}}(t) &=&\left[
\begin{array}{cc}
-1-i & -2i \\
2i & -1+i%
\end{array}%
\right] \breve{a}(t) \\
&&\hspace{-3mm}-\left[
\begin{array}{cc}
K_{-} & -K_{+} \\
-K_{+} & K_{-}%
\end{array}%
\right] \breve{a}_{K}(t)-\left[
\begin{array}{cc}
\sqrt{2} & 0  \\
0 & \sqrt{2}%
\end{array}%
\right] \breve{b}_{in}(t), \\
\dot{\breve{a}}_{K}(t) &=&-\left[
\begin{array}{cc}
2i & \sqrt{2}i \\
-\sqrt{2}i & -2i%
\end{array}%
\right] \breve{a}_{K}(t)+\left[
\begin{array}{cc}
K_{-} & K_{+} \\
K_{+} & K_{-}%
\end{array}%
\right] \breve{a}(t).
\end{eqnarray*}
Define a performance variable $\breve{z}=\breve{a}(t)$. Define system matrices%
\[
A_{cl}=\left[
\begin{array}{cccc}
-1-i & -2i & -K_{-} & K_{+} \\
2i & -1+i & K_{+} & -K_{-} \\
K_{-} & K_{+} & -2i & -\sqrt{2}i \\
K_{+} & K_{-} & \sqrt{2}i & 2i%
\end{array}%
\right] ,
$$
$$
B_{cl}=\left[
\begin{array}{cc}
-\sqrt{2} & 0 \\
0 & -\sqrt{2} \\
0 & 0 \\
0 & 0%
\end{array}%
\right] , \, \, C_{cl}=\left[
\begin{array}{cccc}
1 & 0 & 0 & 0 \\
0 & 1 & 0 & 0%
\end{array}%
\right] .
\]
As in section \ref{sec:analysis-lqg}, the LQG control problem is to minimize, with respect to the direct coupling parameters $K_-$, $K_+$, the LQG performance criterion $
\mathfrak{J}_{\infty}=\mbox{Tr}\left\{ C_{cl}P_{LQG}C_{cl}^{\dagger }\right\}$,
where $P_{LQG}$ is the unique solution of the Lyapunov equation
\begin{equation}
A_{cl}P_{LQG}+P_{LQG}A_{cl}^{\dagger }+\frac{1}{2}B_{cl}B_{cl}^{\dagger }=0,  \label{Lypa}
\end{equation}%
subject to the constraint that $A_{cl}$ is Hurwitz.\footnote{Note that the plant itself is not stable, hence $\mathfrak{J}_{\infty}=\infty $ in the absence of stabilizing direct coupling.}

There are many sets of $\left( K_{-},K_{+}\right) $ such that $A_{cl}$ is not Hurwitz. Consequently, the Matlab function `fminsearch' does not work well. Now if  $K_{-},K_{+}\in \lbrack -15,15]$, $A_{cl}$ is Hurwitz and a simple search finds that $ \mathfrak{J}_{\infty} =1.0406$ at $K_{-}=-15$ and $K_{+}=11.1000$. Extending the ranges of $K_{-}$
and $K_{+}$ will make $\mathfrak{J}_{\infty}$ decrease, but very slightly.

\begin{remark}{\rm
If $\Omega _{+}^{(2)}=3$, no direct coupling of the form (\ref{eq:direct-1}) with real
$K_{-}$ and $K_{+}$ can yield a stable composite system. As a result, we
have to allow $K_{-}$ and $K_{+}$ to be complex numbers. Joint effort of
both direct and indirect couplings may do a better job.} $\Box$
\end{remark}

\subsubsection{The General Case}

The simple example in section \ref{syn:LQG_simple_example} tells us that stability of closed-loop systems poses a big obstacle to quantum LQG control via general-purpose optimization algorithms. In this section a more
systematic algorithm is proposed.

Given a quantum plant $P$ of the form (\ref{plant_optical}) \textit{with the absence of} the quantum signal $\breve{w}(t)$ and noise input $\breve{b}_{v}(t)$, a fully quantum controller (\ref{controller_optical2}) is to be
designed to minimize the infinite-horizon LQG cost defined in section \ref{sec:analysis-lqg}.

For ease of presentation, denote the following matrices
$$
A_{cl}=\left[
\begin{array}{cc}
A & B_{u}C_{K}+B_{12} \\
B_{K}C+B_{21} & A_{K}%
\end{array}%
\right] ,
$$
$$G_{cl}=\left[
\begin{array}{ccc}
B_{f} & B_{u}B_{K0} & 0 \\
B_{K}D_{f} & B_{K1} & B_{K2}%
\end{array}%
\right],
$$
$$C_{cl}=\left[
\begin{array}{cc}
C_{p} & D_{u}C_{K}%
\end{array}%
\right] .
$$%
Clearly, $A_{cl}$ is the the $A$-matrix of the closed-loop system (\ref{syn_cls}). If $A_{cl}$ is Hurwitz, a positive definite matrix $P_{LQG}$ is the unique solution to the following Lyapunov equation%
\begin{equation}
A_{cl}P_{LQG}+P_{LQG}A_{cl}^{\dagger }+\frac{1}{2}G_{cl}G_{cl}^{\dagger
}=0.  \label{Lypa_LQG}
\end{equation}
Following the development in section \ref{sec:analysis-lqg}, the LQG control objective is to design a controller (\ref{controller_optical2}) such that the performance index $\mathfrak{J}_{\infty}=\mbox{Tr}\left\{ C_{cl}P_{LQG}C_{cl}^{\dagger }\right\} $ is minimized, subject to equation (\ref{Lypa_LQG}) and the physical
realizability condition (\ref{eq:direct-5})-(\ref{eq:direct-7}) and (\ref{LQG:physi-reali})-(\ref{LQG:physi-reali_c}).

In \cite{NJP09} an indirect coupling is designed to address the preceding LQG control problem, where it is shown that this problem turns out to be more challenging than the $H^{\infty}$ quantum control because the nice property of separation of control and physical realizability does not hold any more. Therefore, a numerical procedure based on semidefinite programming is proposed to design the indirect coupling. In order to design both direct and indirect couplings, a multi-step optimization algorithm like that in section \ref{algorithm} can be developed. The effectiveness of such an optimization algorithm is illustrated by the following example.

\subsubsection{Example}\label{LQG:ex}

Consider the quantum plant in (\ref{example}) with parameters $\Delta =0.1$ and  $k_{1}=k_{2}=k_{3}=0.01$. This plant can be used to model an atom trapped between two mirrors of a three mirror cavity in the strong coupling limit so that the cavity dynamics can be adiabatically eliminated (see, e.g., \cite{DJ99}). To guarantee the finiteness of the LQG performance, assume $\tilde{w}_1 (t) \equiv 0 $ in (\ref{example}).  According to \cite[Theorem 3.4]{JNP08}, $\tilde{y}(t)$ is the field output quadrature corresponding to the field input quadrature $\tilde{b}_{in,1}(t)$. Note that the quantum plant is marginally stable. A stabilizing indirect coupling is designed in \cite[sec. 8]%
{NJP09} with parameters
$$
\tilde{A}_{K}=\left[
\begin{array}{cc}
0.0257 & -0.3789 \\
0.0666 & -0.2125%
\end{array}%
\right],\tilde{B}_{K1}=\left[
\begin{array}{cc}
0.1126 & -0.5992 \\
0.1504 & -0.1283%
\end{array}%
\right],
$$%
$$
\tilde{B}_{K2}=10^{-10}\left[
\begin{array}{cc}
-0.2721 & 0.0272 \\
-0.1096 & 0.0601%
\end{array}%
\right], ~~\tilde{B}_{K0}=I
$$
$$
\tilde{B}_{K}=\left[
\begin{array}{cc}
1.0297 & -0.1974 \\
0.8255 & -0.0503%
\end{array}%
\right],\tilde{C}_{K}=\left[
\begin{array}{cc}
0.1283 & -0.5992 \\
0.1504 & -0.1126%
\end{array}%
\right].
$$%
The resulting LQG cost $\mathfrak{J}_{\infty}$ is $4.1793$, which greatly outperforms the performance ($\mathfrak{J}_{\infty}=5.4$) achieved by a classical controller modulating a light beam to drive the fully quantum plant \cite{NJP09}. Next we add direct coupling to improve further. By choosing
$$
\tilde{B}_{12 }=10^{-3}\left[
\begin{array}{cc}
1.2 & -9 \\
0.72 & 0.36%
\end{array}\right],\tilde{B}_{21 }=10^{-3}\left[
\begin{array}{cc}
-0.36 & -9 \\
0.72  & -1.2%
\end{array}%
\right] .
$$%
Then it can be verified readily that $\mathfrak{J}_{\infty}=4.000049633093338$. The combination of  direct and indirect
couplings offers an improvement in performance  compared to the controller design in \cite{NJP09}  involving only indirect coupling.

\section{Conclusion} \label{sec:conclusion}

In this paper, we have investigated the influences and uses of indirect and direct couplings in coherent feedback
control of linear quantum stochastic feedback systems. In particular, we have shown that the uses of direct coupling can have beneficial performance consequences, and that the design of direct couplings may be achieved  in a systematic, optimization-based approach. The results of this paper will help to build an integrated, first-principles methodology for coherent quantum control. Future work will include further practical application of the synthesis method of direct couplings in the field of quantum optics.

\section*{Acknowledgment}

The authors wish to thank H. Nurdin for his helpful discussions. The first author would like to thank the second author who introduces him to the field of quantum control.



\begin{thebibliography}{99}


\bibitem{AV73}
B.D.O.~Anderson and S.~Vongpanitlerd,
\newblock {\em Network Analysis and Synthesis}.
\newblock Prentice-Hall, Englewood Cliffs, NJ (also, Dover edition, 2006), 1973.

\bibitem{Belavkin83}
 V.P.~Belavkin,
\newblock ``On the theory of controlling observable quantum systems,''
\newblock {\em Automation and Remote Control}, vol. 44, no. 2, pp. 178-188, 1983.


\bibitem{Belavkin92}
V.P.~Belavkin,
\newblock ``Quantum stochastic calculus and quantum nonlinear filtering,''
\newblock {\em J. Multivariate Anal.}, vol. 42, no. 2, pp. 171-201. 1992


\bibitem{BNM09}
V.P.~Belavkin, A.~Negretti, and K.~Molmer,
\newblock ``Dynamical programming of continuously observed quantum systems,''
\newblock {\em Phys. Rev. A}, vol. 79, 022123 2009.




\bibitem{BvHJ07}
L.~Bouten, R.~Van Handel, and M.R.~James,
\newblock ``An introduction to quantum filtering,''
\newblock  {\em SIAM J. Control and Optimization}, vol. 46, no. 6, pp. 2199--2241, 2007.


\bibitem{BvHJ09}
L.~Bouten, R.~van Handel, and M.R.~James,
\newblock ``A discrete invitation to quantum filtering and feedback control,''
\newblock {\em SIAM Review}, 51, pp. 239-316, 2009.


\bibitem{BMR09}
S.~Bonnabel, M.~Mirrahimi, and P.~Rouchon,
\newblock ``Observer-based Hamiltonian identification for quantum systems,''
\newblock {\em Automatica}, vol. 45, pp. 1144-1155, 2009.


\bibitem{BR04}
H.A.~Bachor and T.C.~Ralph
\newblock {\em A Guide to Experiments in Quantum Optics, 2nd ed.},
\newblock Weinheim, Germany: Wiley-VCH, 2004.




\bibitem{CLG08}
B.A.~Chase, A.J.~Landahl, and J.M.~Geremia,
\newblock ``Efficient feedback controllers for continuous-time quantum error correction,''
\newblock {\em Phys. Rev. A}, vol. 77, 032304, 2008.


\bibitem{CMG07}
R.L.~Cook, P.J.~Martin, and J.M.~Geremia,
\newblock ``Optical coherent state discrimination using a closed-loop quantum measurement,''
\newblock {\em Nature},  vol. 446, pp. 774-777, 2007.


\bibitem{CWJ08}
J.~Combes, H.M.~Wiseman, and K.~Jacobs,
\newblock ``Rapid measurement of quantum systems using feedback control,''
\newblock {\em Phys. Rev. Lett.}, vol. 100, 160503, 2008.


\bibitem{CRH08}
A.R.R.~Carvalho, A.J.S.~Reid, and J.J.~Hope,
\newblock ``Controlling entanglement by direct quantum feedback,''
\newblock {\em Phys. Rev. A}, vol. 78, 012334, 2008.


\bibitem{DJ99}
A.C.~Doherty and K.~Jacobs,
\newblock ``Feedback-control of quantum systems using continuous state-estimation,''
\newblock {\em Phys. Rev. A}, vol. 60, pp. 2700-2711, 1999.


\bibitem{GZ04}
 C.~Gardiner and P. Zoller,
 \newblock {\em Quantum Noise, 3rd ed.},
 \newblock  Springer, Berlin, 2004.



\bibitem{GJ08}
J.E.~Gough and M.R.~James,
\newblock ``The series product and its application to quantum feedforward and feedback networks,''
\newblock  {\em IEEE Trans. Automatic Control}, vol. 54, no.11, 2530-2544, 2009.


\bibitem{GJN10}
J.E.~Gough, M.R.~James and H.I.~Nurdin,
\newblock ``Squeezing components in linear quantum feedback networks,''
\newblock {\em Phys. Rev. A}, vol. 81, 023804, 2010.


 \bibitem{Gough08}
 J.E.~Gough,
\newblock ``Construction of bilinear control Hamiltonians using the series product and quantum feedback,''
\newblock {\em Phys. Rev. A}, vol. 78, 052311, 2008.






\bibitem{GJ09}
J.E.~Gough and M.R.~James,
\newblock ``Quantum feedback networks:  Hamiltonian formulation,''
\newblock {\it Commun. Math. Phys.}, vol. 287, pp. 1109-1132, 2009.

\bibitem{JG10}
M.R.~James and J.E.~Gough.
\newblock ``Quantum dissipative systems and feedback control design by interconnection,''
\newblock {\em IEEE Trans. Automatic Control}, vol. 55, no. 8, 1806-1821, 2010.



 \bibitem{JNP08}
 M.R.~James, H.I.~Nurdin, and I.R.~Petersen,
 \newblock ``$H^{\infty }$ control of linear quantum stochastic systems,''
 \newblock {\em IEEE Trans. Automat. Control, vol.  53, pp. 1787-1803, 2008.}


\bibitem{JS08}
K.~Jacobs and A.~Shabani,
\newblock ``Quantum feedback control: how to use verification theorems and viscosity solutions to find optimal protocols,''
\newblock {\em Contemporary Physics}, vol. 49, pp. 435-448, 2008.

\bibitem{KNPM10}
J.~Kerckhoff, H.I.~Nurdin, D.~Pavlichin, and H.~Mabuchi,
\newblock ``Designing quantum memories with embedded control: photonic circuits for autonomous quantum error correction,''
\newblock{\em Phys. Rev. Lett.,}  vol.105, 040502, 2010.

\bibitem{KY09}
K.~Kashima and N.~Yamamoto,
\newblock ``Control of quantum systems despite feedback delay,''
\newblock {\em IEEE Trans. Automatic Control}, vol. 54, pp. 876-881, 2009.

\bibitem{Lloyd00}
S.~Lloyd,
\newblock ``Coherent quantum feedback,''
\newblock {\em Phys. Rev. A}, vol. 62, 022108, 2000.




\bibitem{Mabuchi08}
H.~Mabuchi,
\newblock ``Coherent-feedback quantum control with a dynamic compensator,''
\newblock {\em Phys. Rev. A}, vol. 78, 032323, 2008.



\bibitem{MK05}
H.~Mabuchi and N.~Khaneja,
\newblock ``Principles and applications of control in quantum systems,''
\newblock {\em I. J. Robust Nonlinear Control}, vol. 15, no. 15, pp. 647--667, 2005.


 \bibitem{MP09}
 A.~Maalouf and I.R.~Petersen.
 \newblock ``Coherent $H_{\infty}$ control for a class of linear complex quantum systems,''
 \newblock  In {\em Proc. American Control Conference}, pages 1472-1479, 2009.



\bibitem{MvH07}
M.~Mirrahimi and R.~van Handel,
\newblock ``Stabilizing feedback controls for quantum system,''
\newblock {\em SIAM J. control and Optimization,} vol. 46, no. 2, pp. 445-467, 2007.



\bibitem{NJP09}
H.I.~Nurdin, M.~R.~James, and I.R.~Petersen,
\newblock ``Coherent quantum LQG control,''
\newblock  {\em Automatica}, vol. 45, pp. 1837-1846, 2009.


\bibitem{NJD09}
H.I.~Nurdin, M.R.~James, and A.C.~Doherty.
\newblock ``Network synthesis of linear dynamical quantum stochastic systems,''
\newblock {\em SIAM J. Control and Optim.}, vol. 48, pp. 2686-2718, 2009.


\bibitem{Pa92}
K.~Parthasarathy.
\newblock {\em An Introduction to Quantum Stochastic Calculus,}
\newblock Berlin, Germany: Birkhauser, 1992.



\bibitem{SGCr97}
C.~Scherer, P.~Gahinet, and M.~Chilali,
\newblock ``Multiobjective output-feedback control via LMI optimization,''
\newblock{\em  IEEE Trans. Automatic Control} vol. 42, no. 7, pp. 896-911, 1997.

\bibitem{SP09}
S.~Shaiju and I.~Petersen.
\newblock ``On the physical realizability of general linear quantum stochastic
  differential equations with complex coefficients,''
\newblock In {\em Proc. 48th IEEE CDC}, pages 1422--1427, 2009.



\bibitem{SHCH09}
S.S.~Szigeti, M.R.~Hush, A.R.R.~Carvalho, and J.J.~Hope,
\newblock ``Continuous measurement feedback control of a Bose-Einstein condensate using phase-contrast imaging,''
\newblock {\em Phys. Rev. A}, vol. vol. 80, 013614, 2009.



\bibitem{SM06}
J.F.~Sherson and K.~Molmer,
\newblock ``Polarization squeezing by optical Faraday rotation,''
\newblock {\em Phys. Rev. Lett.}, vol. 97, 143602, 2006.


\bibitem{SSM08}
G.~Sarma, A.~Silberfarb, and H.~Mabuchi,
\newblock ``Quantum stochastic calculus approach to modeling double-pass atom-field coupling,''
\newblock {\em Phys. Rev. A}, vol. 78, 025801, 2008.


\bibitem{SvHM04}
J.K.~Stockton, R.~Van Handel, and H.~Mabuchi,
\newblock ``Deterministic Dicke-state preparation with continuous measurement and control,''
\newblock {\em Phys. Rev. A},  vol. 70, 022106, 2004.

\bibitem{TMW02}
L.~Thomsen, S.~Mancini, and H.M.~Wiseman,
\newblock ``Continuous quantum nondemolition feedback and unconditional atomic spin squeezing,''
\newblock {\em Phys. Rev. A}, vol. 65, 061801, 2002.




\bibitem{vdS96}
A.~van der Schaft,
\newblock ``$L_2$ Gain and Passivity Techniques in Nonlinear Control'',
\newblock Springer-Verlag New York, Inc. Secaucus, NJ, USA, 1996.


\bibitem{vHSM05}
R.~van Handel, J.K.~Stockton, and H.~Mabuchi,
\newblock ``Feedback control of quantum state reduction,''
\newblock {\em IEEE Trans. Automatic Control}, vol. 50, pp. 768-780, 2005.


\bibitem{vHSM05b}
R.~van Handel, J.K.~Stockton, and H.~Mabuchi,
\newblock ``Modeling and feedback control design for quantum state preparation,''
\newblock {\em J. Opt. B: Quantum Semiclass. Opt.}, vol. 7, S179, 2005.


\bibitem{Wil07}
J.C.~Willems,
\newblock ``The behavioral approach to open and interconnected systems,''
\newblock {\em IEEE Control Systems Magazine}, vol. 27, no. 6, pp. 46-99, 2007.



\bibitem{WT02}
J.C.~Willems and H.L.~Trentelman,
\newblock``Synthesis of dissipative systems using quadratic differential forms: Part I,''
\newblock {\em IEEE Trans. Automatic Control}, vol. 47, no. 1, pp. 53-69, 2002.


\bibitem{WM93}
H.M.~Wiseman and G.J.~Milburn,
\newblock ``Quantum theory of optical feedback via homodyne detection,''
\newblock {\em Phys. Rev. Lett.}, vol.  70, pp. 548-551, 1993.


\bibitem{WM94}
H.M.~Wiseman and G.J.~Milburn,
\newblock ``All-optical versus electro-optical quantum-limited feedback,''
\newblock {\em Phys. Rev. A}, vol. 49, no. 5, pp. 4110-4125, 1994.

\bibitem{WMW02}
H.M.~Wiseman, S.~Mancini, and J.~Wang,
\newblock ``Bayesian feedback versus Markovian feedback in a two-level atom,''
\newblock {\em Phys. Rev. A}, vol. 66, 013807, 2002.



\bibitem{WM09}
H.W.~Wiseman and G.J.~Milburn.
\newblock {\em Quantum Measurement and Control}.
\newblock Cambridge University Press, Cambridge, UK, 2009.



\bibitem{YB09}
N.~Yamamoto and L.~Bouten,
\newblock ``Quantum risk-sensitive estimation and robustness,''
\newblock {\em IEEE Trans. Automatic Control}, vol. 54, no. 1, pp. 92-107, 2009.


\bibitem{YJ09}
M.~Yanagisawa and M.R.~James,
\newblock ``Atom-laser coherence via multi-loop feedback control,''
\newblock {\em Phys. Rev. A}, vol. 79, 023620, 2009.

\bibitem{YK03a}
M.~Yanagisawa and H.~Kimura.
\newblock ``Transfer function approach to quantum control-part I: Dynamics of
  quantum feedback systems,''
\newblock {\em IEEE Trans. Automatic Control}, vol. 48, pp. 2107-2120, 2003.





\bibitem{Car93}
H.J.~Carmichael,
\newblock ``Quantum trajectory theory for cascaded open systems,''
\newblock {\em Phys. Rev. Lett.}, vol. 70, 2273, 1993.


\bibitem{DDJW06}
C.~D'Helon, A.C.~Doherty, M.R.~James, and S.D.~Wilson,
\newblock ``Quantum risk-sensitive control,''
\newblock In {\em Proc. 45th IEEE CDC}, pages 3132-3137, 2006.

\bibitem{DHJMT00}
A.C.~Doherty, S.~Habib, K.~Jacobs, H.~Mabuchi, and S.M.~Tan,
\newblock ``Quantum feedback control and classical control theory,''
\newblock {\em Phys. Rev. A}, vol. 62, 012105, 2000.

\bibitem{Gar93}
C.W.~Gardiner,
\newblock ``Driving a quantum system with the output field from another driven quantum system,''
\newblock {\em Phys. Rev. Lett.}, vol. 70, 2269, 1993.

\bibitem{WD05}
H.M.~Wiseman and A.C.~Doherty,
\newblock ``Optimal unravellings for feedback control in linear quantum systems,''
\newblock {\em Phys. Rev. Lett.},  vol. 94, 070405, 2005.


\bibitem{WM08}
D.F.~Walls and G.J.~Milburn,
\newblock {\em Quantum Optics, 2nd edition},
\newblock Springer, 2008.

\bibitem{YD84}
B.~Yurke and J.S.~Denker,
\newblock ``Quantum network theory,''
\newblock {\em Phys. Rev. A}, vol. 29,  pp. 1419-1437, 1984.

\end{thebibliography}
\end{document}